\documentclass[aps,prl,floatfix,twocolumn,showpacs,preprintnumbers,amsmath,amssymb,superscriptaddress]{revtex4-2}
\usepackage{amsmath}
\usepackage{amssymb}
\usepackage{graphicx}
\usepackage{hyperref}
\usepackage{bm}
\usepackage{epstopdf}
\usepackage{epsfig}
\usepackage{xcolor}

\begin{document}
\title{The birefringent spin-laser as a system of coupled harmonic oscillators}
\author{Velimir Labinac}
\affiliation{Faculty of Physics, University of Rijeka, Rijeka 51000, Croatia}
\author{Jiayu David Cao}
\affiliation{Department of Physics, University at Buffalo, State University of New York, Buffalo, NY 14260, USA}
\author{Gaofeng Xu}
\affiliation{Department of Physics, Hangzhou Dianzi University, Hangzhou, Zhejiang 310018, China}
\author{Igor \v{Z}uti\'c}
\affiliation{Faculty of Physics, University of Rijeka, Rijeka 51000, Croatia}
\affiliation{Department of Physics, University at Buffalo, State University of New York, Buffalo, NY 14260, USA}
\begin{abstract}
Adding spin-polarized carriers to semiconductor lasers strongly changes their properties and, through the transfer of angular momentum, leads to the emission of circularly polarized light. In such spin-lasers the polarization of the emitted light can be modulated an order of magnitude faster than its intensity in the best conventional lasers. This ultrafast operation in spin-lasers relies on large linear birefringence, usually viewed as detrimental in spin and conventional lasers, which couples the two linearly polarized emission modes. We show that the dynamical properties of birefringent spin-lasers under intensity and polarization modulation are accurately described as coupled harmonic oscillators. Our model agrees with the intensity-equation description which, unlike the common complex field components describing the role of birefringence in laser dynamics, uses simpler real quantities and allows analytical solutions. We further predict unexplored operation regimes and elucidate the difference between the weak and strong coupling in spin-lasers.
\end{abstract}
\maketitle

\vspace{-.2cm}
\section{\label{sec:intro}I. Introduction}
\vspace{-.2cm}

Lasers, in addition to having a key role in many applications, with their highly controllable nonlinear response and
coherence~\cite{Chuang:2009,Coldren:2012,Michalzik:2013},
offer model systems to elucidate connections to other cooperative phenomena.
As the injection or pumping of the laser is increased, there is a transition from incoherent to coherent emitted light that can be described by the Landau theory
of second-order phase transitions~\cite{Degiorgio1976:PT}. A mapping can then be established between lasers and ferromagnetism~\cite{Degiorgio1970:PRA}
or Ginzburg-Landau theory of superconductivity~\cite{Haken:1985}, while the instabilities found in lasers directly resemble instabilities found in electronic devices
and hydrodynamics~\cite{Degiorgio1976:PT}. Studies of lasers have also provided a crucial understanding of cooperative phenomena in nonphysical
systems and established the field of synergetics~\cite{Haken:1985}.

\begin{figure}[h!]
\centering
\includegraphics*[width=8.6cm]{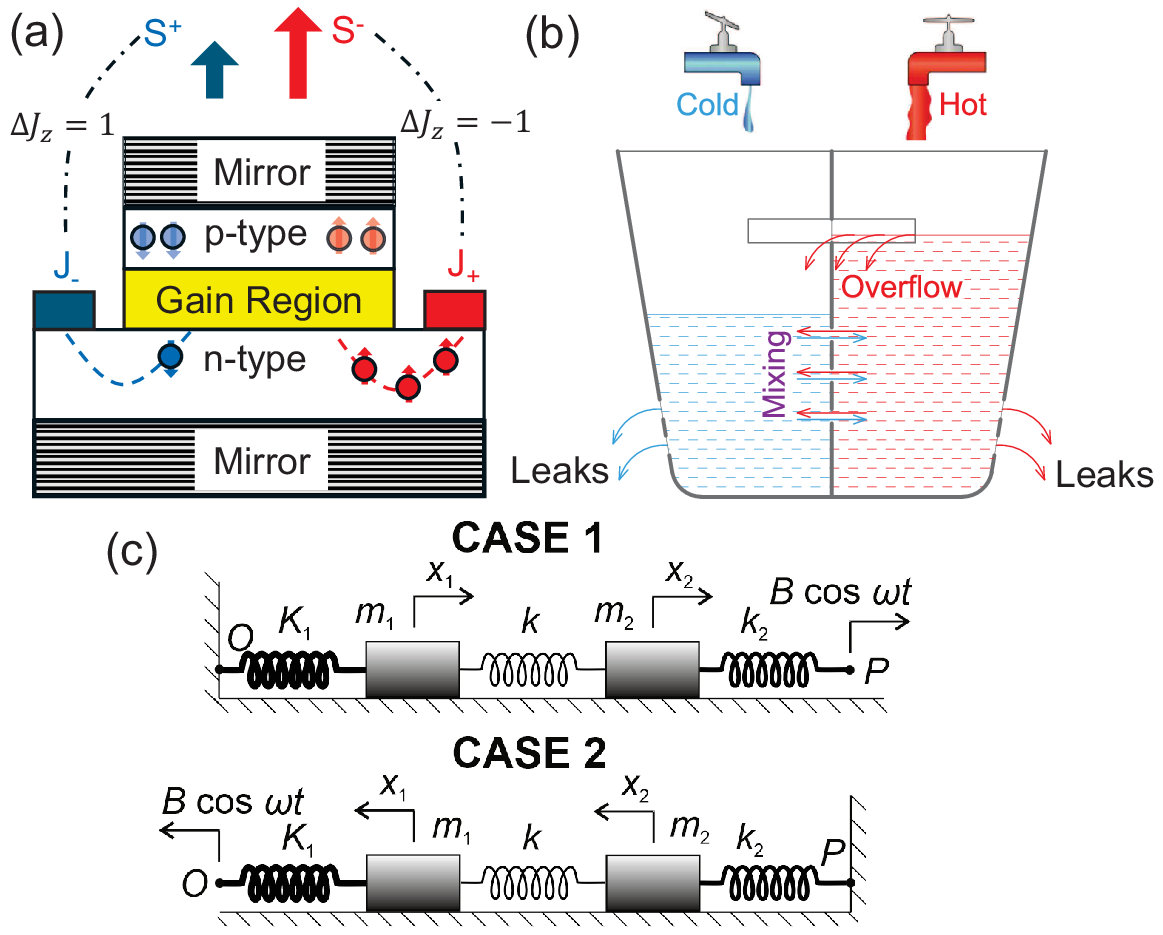}
\vspace{-0.5cm}
\caption{(a) A spin-laser formed by a gain region, $p$- and $n$-type semiconductors, and two mirrors, 
with an unequal injection of different spins ($J_-<J_+$). Circularly polarized emission with photon densities $S^+<S^-$ 
satisfies the optical selection rules of changing
the projection of the total angular momentum $\Delta J_z =\pm 1$.
(b) Spin-bucket model. The two halves denote two spin populations (hot and cold water), separately filled.
An overfilling bucket depicts the lower 
lasing threshold. The imperfect partition mixes the two populations.
(c) Coupled harmonic oscillator model for intensity (CASE 1) and polarization modulation (CASE 2), having masses $m_1$ and $m_2$, with displacements $x_1$ and $x_2$, spring 
constants $K_1 \gg k_2 \gg k$,  a harmonic force displacement of the angular frequency $\omega$ and amplitude $B$.
}
\label{fig:cho1}
\vspace{-0.3cm}
\end{figure}

These prior studies have  
focused on lasers neglecting 
the spin degrees of freedom. 
While both carriers and photons have spin, without 
spin imbalance, just like in 
electronics, this spin can be ignored in conventional lasers.
However, by injecting spin-polarized carriers into semiconductors lasers, as depicted in Fig.~\ref{fig:cho1}(a), such spin imbalance can be generated~\cite{Zutic2020:SSC,Gerhardt2012:AOT}.
Due to the conservation of angular momentum the carrier spin is transferred to photons and it controls the emission of circularly
polarized light~\cite{Hallstein1997:PRB,Zutic2004:RMP}. 
This spin-encoded information can
travel much faster and farther 
than with electron spin,
limited to nanoseconds and microns. 
With the resulting coupled dynamics of carriers, spin,
and the polarization of light, it is then possible to revisit many questions about the connection between 
spin-lasers and other systems~\cite{Lee2012:PRB}.

Rather than exploring how spin-lasers could elucidate other cooperative phenomena, here, we examine
a complementary connection with simple mechanical models which allow us to further understand the operation of these lasers.
Already, a spin-bucket model in Fig.~\ref{fig:cho1}(b), where adding hot (cold) water denotes the injection of spin-up (spin-down) carriers,
suggests a different modulation of lasers, not only by the changing flow but also by the temperature of the added water, representing intensity modulation (IM) and polarization modulation (PM), respectively~\cite{Lee2010:APL,Lee2012:PRB}. 

However, for the dynamical operation of spin-lasers which can be dominated by their anisotropy of the refractive index--birefringence, we show that
a model of coupled harmonic oscillators, depicted in Fig.~\ref{fig:cho1}(c), allows us to further understand the operation of these lasers.
Considering experimental advances in spin-lasers~\cite{Rudolph2005:APL,Gerhardt2006:EL,Holub2007:PRL,Holub2007:PRL,Hovel2008:APL,Basu2009:PRL,%
Saha2010:PRB,Iba2011:APL,Frougier2013:APL,Frougier2015:OE,Cheng2014:NN,Hopfner2014:APL},
which support ultrafast operation and a reduced power consumption~\cite{Zutic2020:SSC,Lindemann2019:N,Yokota2018:APL},
such mechanical models could stimulate progress toward related applications, from spin-encoded information transfer and high-performance interconnects~\cite{Lindemann2019:N}  to neuromorphic computing~\cite{Harkhoe2021:AS}.

We focus  
on the common semiconductor vertical cavity surface emitting lasers (VCSELs)~\cite{Michalzik:2013}.
Without spin-polarized carriers, they  
emit linearly polarized light, 
randomly oriented in the plane of the active (gain) region or  
determined by the two orthogonal directions associated with crystalline orientation and
the anisotropies of the resonant cavity~\cite{Michalzik:2013,SanMiguel1995:PRA}, with photon densities $S_x$ and $S_y$. 
As shown in Fig.~\ref{fig:cho1}(a), this operation is changed when
the spin-polarized carriers are injected, with spin-up/down injection $J_+\neq J_-$, from magnetic
contacts or, alternatively,
using circularly polarized light~\cite{Zutic2020:SSC}. The spin transport is dominated by electrons (bright) since the spin imbalance of holes (pale) is quickly lost with a stronger spin-orbit coupling and they have a much shorter spin-relaxation
time~\cite{Zutic2004:RMP,Fabian2007:APS}. Through the transfer of angular momentum, the spin injection is detected as circularly polarized light. The photon densities of positive and negative helicity $S^+$ and $S^-$ are inequivalent.

The Lorentz model of an atom as a classical damped driven harmonic oscillator~\cite{Chuang:2009} already reproduces many optical properties of semiconductors and can describe dynamical properties of the stimulated emission in conventional lasers~\cite{Siegman:1986,Lee2012:PRB}.
Since we seek to 
understand the polarization properties of the emitted light, one can expect that the coupled harmonic oscillators are needed
to describe the dynamical operation of spin-lasers. A simple manifestation of such coupling is seen  
from the spin-bucket model in Fig.~\ref{fig:cho1}(b):
The two spin populations are coupled through the spin relaxation, as depicted by the holes in the 
partition between two halves of the bucket~\cite{Lee2012:PRB}.
This steady-state coupling 
changes the two  
lasing thresholds and thus also
the dynamical operation of spin-lasers~\cite{Lee2010:APL}.

Models of coupled oscillators continue to elucidate many 
complex phenomena~\cite{Csaba2020:APR,Weisbuch1996:PS,Fruchart2021:N,Hopfield1963:PR,Torma2015:RPP,Segall2017:PRE,Novotny2010:AJP}.
However, even a very simple representation of a dynamical operation of spin-lasers by the oscillations of two masses
coupled by three inequivalent springs and driven by a periodic force poses hundreds of different combinations. 
We analyze the two of them [CASE 1 and CASE 2, in Fig.~\ref{fig:cho1}(c)]
to represent IM and PM, respectively.
Intuitively, the displacements of the two masses will describe the evolution of the two linear polarizations of the emitted light, which are coupled by linear birefringence and, therefore, motivating the presence of the spring connecting the two masses.

To evaluate the relevance of our model of coupled harmonic oscillators, we compare its dynamical results with the calculated intensity equations, known to accurately describe birefringent spin-lasers~\cite{Xu2021:PRB}. Instead of polarization-resolved electric fields with complex amplitudes $E_{x,y}$ for the
optical transitions between the conduction and valence bands, in the intensity equations, it is sufficient to use real-valued photon densities $S_{x,y}=|E_{x,y}|^2$. Similarly, 
for the helicity components, $S^\pm=|E^\pm|^2$, where $E^\pm=(E_x\pm iE_y)/\sqrt{2}$.
The simplicity of the intensity equations 
allows for analytical solutions and provides a direct link with the common rate equations for
both conventional and spin-lasers~\cite{Chuang:2009,Coldren:2012,Michalzik:2013,Rudolph2005:APL,Holub2007:PRL,Gothgen2008:APL,Lee2010:APL,Lee2014:APL,Wasner2015:APL}. The intensity equations are closely related to the spin-flip model (SFM)~\cite{SanMiguel1995:PRA}, introduced to explain the polarization dynamics in conventional VCSELs and later used
for describing
spin-lasers~\cite{SanMiguel1997:IEEE,
Li2010:APL,Gerhardt2011:APL,AlSeyab2011:IEEEPJ,%
Gerhardt2012:AOT,%
Alharthi2015:APL,Lindemann2016:APL,Yokota2018:APL,%
Adams2018:SST,Drong2020:JO,Yokota2021:M,Yokota2021:IEEEPTL,Yokota2025:APLP}.

A recent breakthrough that the spin-orbit torque magnetization switching electrically reverses the helicity of the emitted light from semiconductors at room temperature and zero applied magnetic 
field~\cite{Dainone2024:N,Hiura2024:N} further motivates our goal to understand the modulation of spin-lasers. Unlike prior experiments,
where the modulation of spin-lasers was limited to simple but commercially impractical optical spin injection~\cite{Zutic2020:SSC,Gerhardt2012:AOT}, with this principle,
electrical spin injection and PM in spin-lasers could be used to integrate spintronics, electronics, and photonics. 
The model of coupled harmonic oscillators could elucidate 
unexplored regimes in spin-lasers with the prospect of using them for spin-charge-photon conversion and long-distance transfer of information encoded in the helicity of the emitted light.

Following this introduction, in Sec.~II, we describe our model of
coupled harmonic oscillators. In Sec.~III, we compare this model
with the dynamical operation of spin-lasers calculated from the intensity equations. In Sec.~IV, we investigate the behavior of the quality factor
and the coupling strength. In Sec.~V, we  provide conclusions and note some open questions for future work.

\section{\label{sec:coupled}II. Coupled Oscillators Model}

The insights from models of coupled oscillators are ubiquitous to many areas of physics, from light-matter interaction and phonons to artificial neural networks~\cite{Csaba2020:APR,Weisbuch1996:PS,Fruchart2021:N,Hopfield1963:PR,Torma2015:RPP,Segall2017:PRE,Novotny2010:AJP}.
IM in conventional lasers is already 
accurately described by a
damped driven harmonic oscillator
${\ddot x}+\gamma {\dot x}+\omega^2_0x=(F_0/m) \cos\omega t$,
where $x$ is the displacement, 
$\omega_0=k/m$ is the  natural 
angular frequency of the simple harmonic
oscillator, with mass $m$ and the spring constant $k$,
$\gamma=c/m$ is the damping constant, $c$ is the damping coefficient, $t$ is the time, and $F_0$ is the amplitude
of the driving force.
The normalized displacement amplitude
\begin{equation}
A(\omega)/A(0)=\omega_0^2/\left[(\omega_0^2-\omega^2)^2+\gamma^2\omega^2\right]^{1/2},
\label{eq:HO}
\end{equation}
shares the IM behavior of lasers with resonance 
near $\omega \approx \omega_0$ and a large reduction of $A(\omega)$ for $\omega \gg \omega_0$.
The reduction of  $A(\omega)$ by $-3\ \mathrm{dB}$ compared with $A(0)$ gives a modulation bandwidth, a frequency range over which
substantial signals can be transferred~\cite{Chuang:2009,Lee2012:PRB,Xu2021:PRB,Zutic2020:SSC,Hecht2016:N}.

Despite its wide use for conventional lasers, the model of a single harmonic
oscillator is not sufficient to describe different types of modulation or the coupling between the $x$ and $y$ polarization modes of oscillation in 
spin-lasers. Instead, to elucidate various trends in spin-lasers, we should
seek to model their behavior using coupled harmonic oscillators. We 
expect that the time-dependent displacements $x_1$ and $x_2$ of the masses $m_1$ and $m_2$ 
can be used to model the oscillatory behavior of $S_x$ and $S_y$.

With many possible 
models of coupled harmonic oscillators and their different behaviors, we deliberately choose $K_1\gg k_2$, based on the observed asymmetric $x$ and $y$ modes in spin-lasers~\cite{Zutic2020:SSC,Gerhardt2012:AOT}, due to the anisotropy of the refractive index (birefringence), where 
$n_x \neq n_y$, and the anisotropy of the absorption (dichroism).  

To illustrate these different behaviors 
we simplify CASE 1 and CASE 2 by setting 
$K_1 \rightarrow \infty $ and 
all the other quantities in Fig.~\ref{fig:cho1}(c) to be finite. We examine the motion of  
$m_1$ and $m_2$ when connecting the external harmonic force to the points $P$ or $O$. In CASE 1, $m_1$ does not oscillate at all, and the spring $K_1$ behaves as a stationary rod. In contrast, in CASE 2, the external force moves the rigid spring so both masses oscillate. This is an encouraging agreement for spin-lasers, where we have
preliminary support that IM and PM behavior could be distinguished by the presence of one and two resonance peaks in the dominant $S_y$ mode, respectively, and our notation that capitalizes $K_1$
suggests that its value is larger than for the other spring constants. 
Since the birefringent spin-lasers are expected to operate in a weak coupling regime, we select that the middle spring constant (absent without the birefringence which couples $S_x$ and $S_y$) is the weakest one,  
as indicated in Fig.~\ref{fig:cho1}(c).

By providing next a short overview for the underlying equations describing CASE 1 and 
CASE 2, we also establish a framework to compare their properties with the
modulation response of spin-lasers.
The potential energy for CASE 1 is $ K_1x_1^2/2 + k\left(  x_1-x_2\right)  ^2/2 + k_2\left(  x_2-b\right)  ^2/2$, where 
$K_1 \gg k_2 \gg k$, and $b\left(  t\right)  =B\cos\left(  \omega t\right)$
is displacement of point $P$ from equilibrium by the external force. The equations of motion for $m_1$ and $m_2$ and their displacements 
$x_1\left(  t\right) $ and $x_2\left(
t\right)  $ are 
\begin{align}
\ddot{x}_1 &  =-\frac{K_1}{m_1}x_1-\frac{k}{m_1}\left(x_1%
-x_2\right)  -\frac{c_1}{m_1}\dot{x}_1, 
\label{eq:x1dots} \\ 
\ddot{x}_2 &  =\frac{k}{m_2}\left(x_1-x_2\right)  -\frac{k_2%
}{m_2}\left(x_2-b\right)  -\frac{c_2}{m_2}\dot{x}_2,
\label{eq:x2dots}
\end{align}
where $c_1$ and $c_2$ are the damping coefficients.  For a more
direct comparison with the first-order differential equations describing
spin-lasers, we recast Eqs.~(\ref{eq:x1dots}) and (\ref{eq:x2dots}) by 
introducing a dimensionless parameter $\tau' = t/T$, where $T$ is the time-scale factor, and dimensionless 
variables
\begin{equation}
\eta_1 = x_1/B, \eta_2 = \dot{x}_1/B, \eta_3 = x_2/B, \eta_4 = \dot{x}_2/B,
\label{eq:eta1}
\end{equation}
where the derivatives are 
with respect to $\tau'$. The equations for the two coupled harmonic oscillators become
\begin{align}
\dot{\eta}_1 &  =\eta_2, 
\label{eq:eta1dot} \\
\dot{\eta}_2 &  =-\frac{K_1 + k}{m_1}T^2 \eta_1 - \frac{c_1}{m_1}T \eta_2 + \frac{k}{m_1}T^2 \eta_3, 
\label{eq:eta2dot} \\
\dot{\eta}_3 &  =\eta_4, 
\label{eq:eta3dot} \\
\dot{\eta}_4 &  =\frac{k}{m_2}T^2 \eta_1 - \frac{k+k_2}{m_2} T^2\eta_3%
-\frac{c_2}{m_2}T \eta_4 + \frac{k_2}{m_2 B}T^2 b.
\label{eq:eta4dot}
\end{align}
We use external harmonic force displacement in the form of the real part of complex function $b\left( t \right) = B\exp\left(  -i\omega t \right)$ and seek a particular solution for the system of Eqs.~(\ref{eq:eta1dot})--(\ref{eq:eta4dot}) as
\begin{equation}
\eta_j=u_{j}e^{-i\omega t}, \quad %
j=1,2,3,4.
\label{eq:etaj}
\end{equation}
The amplitudes $u_j\left(
\omega' \right)$ can be expressed as a vector 
$\mathbf{u}\left(
\omega' \right)$ 
and given in the matrix form
\begin{equation}
\mathbf{u}\left(  \omega' \right) = -\left( \bm{\mathsf{U}} + i\omega' \bm{\mathsf{I}}%
\right)^{-1} \mathbf{b}_1,
\label{eq:uomega}
\end{equation}
where $\omega' = \omega T$ is a dimensionless angular frequency, $\bm{\mathsf{I}}$ is the unit matrix
\begin{equation}
\mathbf{b}_1 = \left(0, 0, 0, k_2T^2/m_2\right),
\label{eq:b1}
\end{equation}
\begin{equation}
\bm{\mathsf{U}}=
\begin{pmatrix}
0 & 1 & 0 & 0\\
-\omega_1^2 T^2 & -c_1 T / m_1 & k T^2 / m_1 & 0\\
0 & 0 & 0 & 1\\
k T^2 / m_2 & 0 & -\omega_2^2 T^2 & -c_2T /m_2
\end{pmatrix},
\label{eq:Umatrix}
\end{equation}
with $\omega_1^2 = \left(K_1 + k \right) / m_1$ and $\omega_2^2 = \left( k_2 + k \right) / m_2$. 

By repeating this procedure for CASE 2, where the potential energy is 
$ K_1 \left( b - x_1 \right)^2/2 + k\left(x_1-x_2\right)^2/2 + k_2 x_2^2/2$, we obtain 
Eq.~(\ref{eq:uomega}) but with $\mathbf{b}_{1}$ replaced by 
\begin{equation}
\mathbf{b}_2 = \left(0, K_1T^2 / m_1, 0, 0\right). 
\label{eq:b2}
\end{equation}

The coupled harmonic oscillators in Fig.~\ref{fig:cho1}(c) have two normal modes of vibration with the corresponding eigenfrequencies proportional to the positive imaginary part of the eigenvalues of the matrix $\bm{\mathsf{U}}$. The resonant frequencies are approximately equal to eigenfrequencies of the system because of 
the small damping constants
$c_i/m_{i},i=1,2$. With the damping, the free oscillations which are composed of normal modes, fall off to zero at a rate proportional to the real part of eigenvalues of $\bm{\mathsf{U}}$ and do not contribute significantly to forced oscillations. Here, 
$\bm{\mathsf{U}}$ can be decomposed as the sum of the block-diagonal
\begin{equation}
\bm{\mathsf{H}}_{\mathsf{0}}=
\begin{pmatrix}
0 & 1 & 0 & 0\\
-\omega_{1}^2 T^2 & -c_{1} T / m_1 & 0 & 0\\
0 & 0 & 0 & 1\\
0 & 0 & -\omega_2^2 T^2 & -c_2T /m_2
\end{pmatrix},
\label{eq:Hmatrix}
\end{equation}
and off-diagonal $\bm{\mathsf{V}}$ describing coupling
\begin{equation}
\bm{\mathsf{V}}=%
\begin{pmatrix}
0 & 0 & 0 & 0\\
0 & 0 & k T^2/ m_1 & 0\\
0 & 0 & 0 & 0\\
k T^2 / m_2 & 0 & 0 & 0
\end{pmatrix}.
\label{eq:Vmatrix}
\end{equation}
If $k$ is small, 
$m_1$ and $m_2$ 
are weakly coupled.  

The lowest
order of approximation,  
$\bm{\mathsf{V}} \approx \bm{\mathsf{0}}$ and
$\bm{\mathsf{U}}\approx \bm{\mathsf{H}}_{\mathsf{0}}$ describes 
the two independent harmonic oscillators with well-separated lower and higher angular eigenfrequencies
\begin{eqnarray}  
\omega_L(k=0)=\left[k_2/m_2 - \left(c_2 / 2 m_2 \right)^2\right]^{1/2},
\label{eq:omegaL}
\\
\omega_H(k=0)=\left[K_1/m_1 - \left(c_1 / 2 m_1 \right)^2\right]^{1/2}.
\label{eq:omegaH}
\end{eqnarray} 
While the equations for 
lasers are more complex, this approach will help us to conclude that $S_x$ and $S_y$ oscillations are usually 
also weakly coupled and the off-diagonal matrix for a spin-laser is a small perturbation. 

\section{\label{sec:compar}III. Comparison of dynamical behavior}

\subsection{\label{sec:IE}A. Intensity equations for spin-lasers}

To investigate the relevance of our coupled oscillator model, we compare the trends in its dynamical operation with that for a spin-laser which is characterized with inequivalent spin injection of spin-up/down carriers, which leads to the spin polarization of injected carriers
\begin{equation}
P_J = (J_+-J_-)/(J_++J_-),
\label{eq:PJ}
\end{equation}
with the total injection $J=J_++J_-$. 
Such a spin injection leads to the 
polarization of the emitted light
\begin{equation}
P_{C}=(S^+-S^-)/(S^++S^-),
\label{eq:PC}
\end{equation}
where the total photon density can be expressed in terms
of its helicity or linearly polarized components $S=S^+ + S^-=S_x + S_y$. Remarkably, in highly birefringent spin-lasers, where $\gamma_p$ is the linear birefringence, the changes in $P_C$ can be much faster than the changes in
the intensity of the emitted light~\cite{Lindemann2019:N},
while the spin-encoded information in the helicity of the light can travel much farther and faster than the spin information from the carrier spin, typically limited to microns or nanoseconds~\cite{Zutic2004:RMP}.

To describe such spin-lasers, we use the intensity equations~\cite{Xu2021:PRB} that are closely related to the SFM~\cite{SanMiguel1995:PRA} which is obtained from the Maxwell-Bloch equations~\cite{Hess1996:PRA,Fordos2017:PRA} and frequently used to describe the polarization dynamics in VCSELs. The SFM describes the optical transitions in the quantum-well based gain region between the conduction band, with the projection of the total angular momentum (in units of $\hbar$) $J_z =\pm 1/2$, and the valence band, with $J_z = \pm 3/2$ for heavy holes (split from the light holes~\cite{Zutic2004:RMP}). These transitions lead to the emitted photons with the angular momentum $\pm \hbar$, described in SFM by the corresponding complex amplitudes of electric fields of the positive (negative) helicity. The other two variables in SFM are the total number of carriers $N$ and $n$, the population difference between the spin-up and spin-down electrons. 

The quantities in the SFM equations~\cite{SanMiguel1995:PRA} are usually studied in the dimensionless form, making it important to describe how they are normalized and simplify their relation to other rate-equation description of lasers. Specifically, the
quantities in SFM have been normalized as%
\begin{eqnarray}
E^{\pm} &=& F^{\pm}/\sqrt{S_{2J_T}} , \\
N &=& (N_+ + N_- - N_\mathrm{tran})/(N_T - N_\mathrm{tran}), \\
n &=& (N_- - N_+)/(N_T - N_\mathrm{tran}),
\end{eqnarray}%
where $F^{\pm}$ are the slowly varying amplitudes of the  helicity-resolved components of the electric field, $S_{2J_T}$ is the steady-state light intensity at twice the threshold injection $2J_T$, $N_\pm$ are the numbers of spin-up and spin-down electrons, $N_T$ and $N_\mathrm{tran}$ are the numbers of electrons at the threshold and transparency, respectively.
The injection $J$ has been normalized by the 
threshold injection $J_T$. In the above normalizations, we consider a situation 
for many VCSELs, where the dichroism is much smaller
than the inverse of the photon lifetime~\cite{Xu2021:PRB,Yokota2023:IEICE}. 

Within the intensity equations, which are all expressed in terms of real-valued variables, the complex electric fields for the two
linear modes, with the phase (difference or shift)
$\phi=\phi_x-\phi_y$, 
are replaced by the photon densities, recall Sec.~I. These equations contain several other
important differences from the SFM, and their transparency allows for analytical solutions~\cite{Xu2021:PRB}. 
Since the spin-relaxation
time for holes is much shorter than for electrons, 
we use $\tau_{sp} \ll \tau_{sn} = \tau_s$~\cite{Zutic2004:RMP,Fabian2007:APS}, 
while the SFM assumption $\tau_{sp} = \tau_{sn} = \tau_s$
is not correct for quantum wells. Furthermore, the gain saturation (absent in SFM) is inherent to lasers, and we include it phenomenologically in the intensity equations~\cite{Xu2021:PRB}. 

While $N$ is retained in the intensity equations, since the spin-relaxation
time in 
the gain region is typically much shorter than the carrier recombination time $\tau_s \ll \tau_r = 1/\gamma_r$~\cite{Lindemann2016:APL,Lindemann2019:N}, $n$ can
be adiabatically eliminated ($\dot{n}\approx 0$)~\cite{Xu2021:PRB}. The resulting
equations with 
variables $\left(\phi, S_x, S_y, N \right)$ accurately describe the birefringent 
spin-lasers~\cite{Lindemann2019:N}. They can be written 
in terms of the dimensionless real quantities, while all the frequencies are scaled to $\gamma_r$ and differentiation
expressed with respect to dimensionless time $\tau=\gamma_r t$ as
\begin{align}
\dot{\phi} &  = F_1\left(\phi,S_x,S_y,N\right) , 
\label{eq:phi_dot} \\ 
\dot{S}_x &  = F_2\left(\phi,S_x,S_y,N\right) ,
\label{eq:Sx_dot} \\ 
\dot{S}_y &  = F_3\left( \phi,S_x,S_y,N\right) ,
\label{eq:Sy_dot} \\ 
\dot{N} &  = F_{4}\left(  \phi,S_x,S_y,N\right) ,
\label{eq:N_dot} 
\end{align}
where the functions $F_j = F_j\left(  \phi,S_x,S_y,N\right),\;j = 1,2,3,4$
are given by
\begin{eqnarray}
F_1  &=& -2\gamma_p -\frac{\tau_s}
{2\tau_{\text{ph}}}\left(  J P_J+2N\sqrt{S_x S_y}\sin\phi\right) \nonumber \\
 &&\times \left[
\alpha \sin\phi\left(  \sqrt{S_y/S_x}-\sqrt{S_x/S_y}\right) \right. \nonumber \\
&& \left. +\cos\phi\left(  \sqrt{S_y/S_x}+\sqrt{S_x/S_y}\right)
\right] ,
\label{eq:F1} \\ 
F_2 &=& S_{x}\left[\left(
N-1\right)/\tau_{\text{ph}}  -2\gamma_a-\epsilon_{xy}S_y-\epsilon_{xx}S_x\right] \nonumber \\
&& + \frac{J P_J\tau_s}{\tau_{\text{ph}}} \sqrt{S_x S_y}\left(\alpha \cos\phi
-\sin\phi\right) , 
\label{eq:F2} \\ 
F_3 &=& S_y\left[\left(
N-1\right)/\tau_{\text{ph}}  +2\gamma_a-\epsilon_{yx}S_x-\epsilon_{yy}S_y\right] \nonumber
\\
&& - \frac{J P_J\tau_s}{\tau_{\text{ph}}} \sqrt{S_x S_y}\left(\alpha \cos\phi
+\sin\phi\right) ,
\label{eq:F3} \\ 
F_{4} &=& -N+J - N\left(S_x +S_y\right)  +2\tau_s N S_xS_y.
\label{eq:F4} 
\end{eqnarray}
In Eqs.~(\ref{eq:F1})--(\ref{eq:F4}), $\tau_{\text{ph}}$ is the photon lifetime, $\alpha$ is the linewidth enhancement factor, and $\gamma_a$ is the dichroism, which represents the anisotropy of absorption (or equivalently of optical gain)~\cite{Adams2022:IEEEJQE}. 
As in a simple description of conventional lasers, the gain saturation coefficients are given by $\epsilon_{xx}=\epsilon_{yy}=\epsilon$ and $\epsilon_{xy}=\epsilon_{yx}=0$~\cite{Chuang:2009,Coldren:2012}.
For a more complete description of the gain saturation, we phenomenologically introduce self-saturation terms with coefficients $\epsilon_{xx}$ and $\epsilon_{yy}$ for the $x$ and $y$ modes~\cite{Xu2021:PRB}.

\subsection{\label{sec:IM}B. Intensity modulation}

One of the most attractive properties of lasers is their versatile dynamical response including the use of 
external modulation to attain a large bandwidth. Their modulation response is usually studied within the small-signal analysis, where each of the key quantities ($J$, $S$, $N$, and $P_J$), 
is decomposed into a steady-state and a modulated part. For IM, this means 
\begin{equation}
J =J_0+\delta J\left(t \right), 
\label{eq:IM2} 
\end{equation}
where we assume harmonic modulation, 
$\delta J \left(t \right) = \operatorname{Re} \left[ \delta J\left(\omega \right) e^{-i \omega t} \right]$ 
with 
$\left\vert \delta J\left(t \right) \right\vert \ll 1$ and $P_J=P_{J0}$. 

We recall that, in conventional lasers, where $P_J = 0$, IM is accurately described by a model of a single 
harmonic oscillator and summarized by relating their resonant (relaxation-oscillation) frequency  $f_R=\omega_R^{\text{IM}}/2\pi$ and the resulting modulation bandwidth~\cite{Chuang:2009,Coldren:2012}
\begin{equation}
f_\mathrm{3dB} \approx \sqrt{1+\sqrt{2}} f_R.
\label{eq:3dB}
\end{equation}
The response function for IM~\cite{Chuang:2009}
\begin{equation}
R(\omega)=|\delta S(\omega)/\delta J\left(\omega \right)| 
\label{eq:response}    
\end{equation}
is usually normalized to its $\omega=0$ value 
$R(\omega)/R(0)$, which recovers the form of the dynamical behavior of a single
harmonic oscillator from Eq.~(\ref{eq:HO})
by replacing $\omega_0$ with $\omega_R$. This $\omega_R$ and the damping constant 
$\gamma$ follow 
from the intensity equations given by Eqs.~(\ref{eq:phi_dot})--(\ref{eq:F4}).
For example, assuming $S_x=0$, we can obtain the steady-state value $S_{y0}=J_0/N_0-1$ and $N_0=1-2 \tau_{\text{ph}}\gamma_a + \tau_{\text{ph}} \epsilon_{yy}S_{y0}$
and conclude that the normalized threshold values are
\begin{equation}
J_T=N_T=1-2\tau_{\text{ph}} \gamma_a.
\label{eq:JNT}
\end{equation}
We can then express
\begin{eqnarray}
\omega_R^2&=&(J_0/N_0-1)(N_0/\tau_{\text{ph}}+\epsilon_{yy} J_0/N_0), \label{eq:omegaR} \\
\gamma&=&(J_0/N_0)(1+\epsilon_{yy})-\epsilon_{yy},
\label{eq:gamma}
\end{eqnarray}
and $\omega_R^{\text{IM}} \approx \omega_R$ for $\gamma \ll \omega_R$. Instead, if we assume $S_y=0$, then $\omega_R$, and $\gamma$ would retain the same form but with $\epsilon_{yy}\rightarrow \epsilon_{xx}$.

For IM, the bandwidth can be estimated by~\cite{Chuang:2009} 
\begin{equation}
f_R=(1/2\pi)\sqrt{g_0 S_0/[\tau_{\text{ph}}(1+\epsilon S_0)]},  
\label{eq:fr}
\end{equation}
where $g_0$ is the gain constant,
and $\epsilon$ is the simplified parameterization of the gain saturation, noted in Sec.~IIIA. 
While the common approach is to enhance the bandwidth by increasing $J$ to attain a larger $S$, its drawback is the higher-power consumption and that a finite $\epsilon$ is responsible for the saturation of $S$ as $J$ is increased. 

However, for birefringent spin-lasers, many dynamical features can no longer be
described by a model of a single harmonic oscillator. 
This has three important implications:
(i) One should instead examine the relevance of modeling such lasers as coupled harmonic oscillators. (ii) There are other paths to enhance the bandwidth and a suitable dynamical operation, beyond the guidance suggested from Eq.~(\ref{eq:fr}), including a striking role
of birefringence~\cite{FariaJunior2015:PRB,Lindemann2019:N}. The resulting
large frequency splitting, $\Delta f> 200\ \mathrm{GHz}$ between $x$ and $y$ modes, leads to the polarization oscillations with the resonant frequency, as the beat
frequency between the two orthogonal modes
\begin{equation}
f^\text{PO}_R \approx \gamma_p/\pi,
\label{eq:fr_tilde}
\end{equation}
where $f^\text{PO}_R \gg f_R$.
(iii) The phase evolution 
needs to be carefully examined. Unlike in the single harmonic oscillator where, above  the resonance in the amplitude,  it 
is accompanied by the $\pi$ shift in the phase (more abrupt for $\gamma \ll 1$~\cite{Fowles:2005}), the phase evolution can be very different in coupled harmonic oscillators and birefringent
spin-lasers.

For our small-signal analysis, we choose a stable state with $S_{x} \ll 1$ and $S_{x} \ll S_{y}$ such that the light is almost completely $y$ polarized,  
while the coupling between $S_x$ and $S_y$ through $\gamma_p$ is still strong enough to provide interesting effects in various modulation schemes, including PM as well as for a combined IM and PM.

By recognizing the structure of the dimensionless intensity equations in Eqs.~(\ref{eq:phi_dot})--(\ref{eq:N_dot}), it is convenient to rewrite
them in a compact form
\begin{equation}
\mathbf{\dot{X}}=\mathbf{F}\left(\mathbf{X}\right),
\label{eq:Xdot}
\end{equation}
where $\mathbf{X} = \left(\phi, S_x, S_y, N \right)$ and
$\mathbf{F}\left(  \mathbf{X}\right)  = \left(F_1, F_2, F_3, F_4 \right)$. We seek the solution of Eq.~(\ref{eq:Xdot}) in the form appropriate for the small-signal analysis
\begin{equation}
\mathbf{X} = \mathbf{X}_{0} + \delta \mathbf{X},
\label{eq:X}
\end{equation}
where the components of 
$\delta \mathbf{X}=\left(\delta\phi, \delta S_x,\delta S_y, \delta N \right)$
satisfy $ \left\vert \delta X_j \right\vert \ll 1$. The corresponding system
of the linearized equations obtained from Eq.~(\ref{eq:Xdot}) using the Taylor series 
expansion at the point $\mathbf{X}_0=\left(\phi_0, S_{x0}, S_{y0}, N_0\right)$ has the leading
contribution given by
\begin{equation}
\mathbf{\dot{X}}=\mathbf{F}\left(  \mathbf{X}_{0} \right)
+\bm{\mathsf{M}}\left(  \mathbf{X}_0 \right) \delta\mathbf{X},
\label{eq:Xlin}
\end{equation}
with the elements of the Jacobian matrix $\bm{\mathsf{M}}\left(  \mathbf{X}_0\right)$, evaluated at 
a stable critical point $\mathbf{X}_0$,
given by
\begin{equation}
M_{ij} = \left( \partial F_i/\partial X_j \right)_0.
\label{eq:Mij}
\end{equation}
By applying this procedure to IM from Eq.~(\ref{eq:IM2}), we get
\begin{equation}
\delta \mathbf{\dot{X}}=\bm{\mathsf{M}}\left( \mathbf{X}_0,J_0\right)
\delta \mathbf{X}+\mathbf{G}^{\text{IM}}\left( \mathbf{X}_0,J_0\right)
\delta J,
\label{eq:SSA_IM}
\end{equation}%
where $\mathbf{G}^{\text{IM}} = \left(G^{\text{IM}}_1, G^{\text{IM}}_2, G^{\text{IM}}_3, G^{\text{IM}}_4 \right)$ and
\begin{eqnarray}
G^{\text{IM}}_1 &=& -\frac{\tau_s}{2 \tau_{\text{ph}}}P_{J0}\left[\alpha \sin\phi_0\left(  \sqrt{
S_{y0}/S_{x0}}-\sqrt{S_{x0}/S_{y0}}\right) \right. \nonumber \\
 &&\left. +\cos\phi_0\left(\sqrt{S_{y0}/S_{x0}}+\sqrt{S_{x0}/S_{y0}}\right) \right], 
\label{eq:G1_IM} \\
G^{\text{IM}}_2 &=& \frac{\tau_s}{\tau_{\text{ph}}}P_{J0}\sqrt{S_{x0}S_{y0}}\left(  \alpha \cos\phi_0%
-\sin\phi_0\right), 
\label{eq:G2_IM} \\
G^{\text{IM}}_3 &=& -\frac{\tau_s}{\tau_{\text{ph}}}P_{J0}\sqrt{S_{x0}S_{y0}}\left(  \alpha \cos\phi_0%
+\sin\phi_0\right), 
\label{eq:G3_IM} \\
G^{\text{IM}}_4 &=& 1.
\label{eq:G4_IM}
\end{eqnarray}

For a harmonic IM  the particular solution for Eq.~(\ref{eq:SSA_IM}) has the form
\begin{equation}
\delta \mathbf{X}=\delta \mathbf{X}_0e^{-i\omega t},
\label{eq:deltaX}
\end{equation}%
where $\delta \mathbf{X}_0$ is $\omega$ dependent. 
We obtain 
\begin{equation}
\delta \mathbf{X}_0\left( \widetilde{\omega} \right) =-\left( \bm{\mathsf{M}}+i \widetilde{\omega} %
\bm{\mathsf{I}}\right) ^{-1}\mathbf{G}^{\text{IM}}\delta J_0,
\label{eq:deltaX0}
\end{equation}%
where $\bm{\mathsf{I}}$ is the unit matrix, $\widetilde{\omega} = \omega / \gamma_r$ is the dimensionless angular frequency, and $\delta J\left(\omega \right)$ is assumed frequency independent, $\delta J\left(0 \right) = \delta J_{0}$. For a small modulation
$|\delta J_0| \ll 1$, we can obtain IM results  
that are independent of the magnitude of $\delta J_0$. 

With this framework, we can compare the modulation response for CASE 1 of coupled harmonic oscillators with the corresponding results for spin-lasers.
We study the oscillation of $m_2$, described by the displacement $x_2$ in Fig.~\ref{fig:cho1}(c) or, equivalently, by the dimensionless amplitude $u_3(\omega')$ 
in Eq.~(\ref{eq:etaj}), a part of the matrix form in Eq.~(\ref{eq:uomega}), and compare it with the evolution of $S_y$ from IM in spin-lasers described by 
$\delta S_{y0}$ in Eq.~(\ref{eq:deltaX0}). 

\begin{figure}[t]
\centering
\includegraphics*[width=8.6cm]{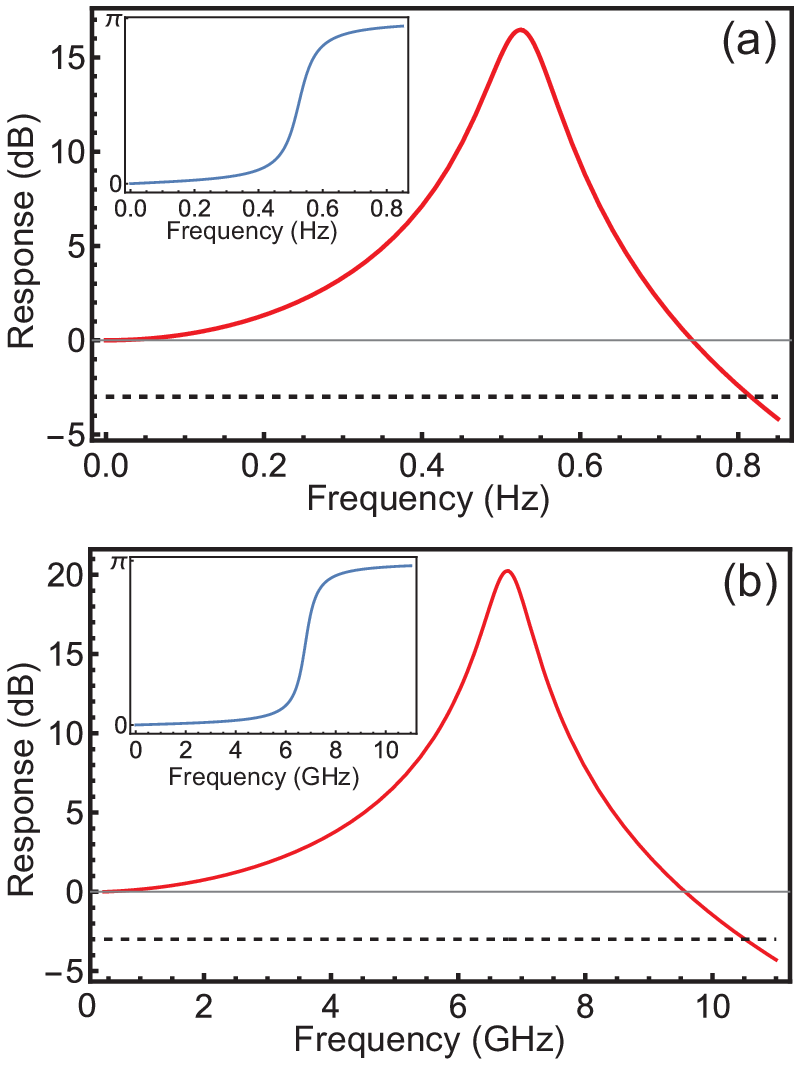}
\vspace{-0.5cm}
\caption{(a) Normalized response function for CASE 1 of the coupled harmonic oscillators from Fig.~\ref{fig:cho1}(c), describing the frequency-dependent displacement $x_2(f)$. 
Dashed line: $-3\ \mathrm{dB}$ response marks the modulation bandwidth. Inset: Phase shift 
evolution $\phi_M(f)$. 
The parameters are spring constants $K_1 = 100\,k = 10\,k_2 = 100\ \mathrm{g/s^2}$, masses $m_1 = 2\,m_2 = 2\ \mathrm{g}$, damping coefficients $c_1 = 0.7\ \mathrm{g/s}$, $c_2 = 0.5\ \mathrm{g/s}$, $B = 3\ \mathrm{cm}$, and the 
scale factor $T = 1\ \mathrm{s}$.
(b) IM response function of the spin-laser for photon density $S_{y}$ with the marked $-3\ \mathrm{dB}$ response and the phase 
shift evolution (inset). The parameters, 
defined in Sec.~IIIA, are $\gamma_p = 75 \pi\ \mathrm{GHz}$, $\tau_\mathrm{ph} = 1.67\ \mathrm{ps}$, $J_0 = 4.0 J_T$, $P_{J0} = 0.04$, $\gamma_s = 1/\tau_s = 300\ \mathrm{GHz}$, $\gamma_a = 10\ \mathrm{GHz}$, $ \epsilon_{xx} =\epsilon_{yy} = 0$, $\epsilon_{xy} = \epsilon_{yx} = \tau_s/\tau_\text{ph}$, and $\alpha = 2$.
}
\label{fig:cho2}
\vspace{-0.3cm}
\end{figure}

To obtain the information of the modulation response and the corresponding bandwidth, not just for IM but also for other modulation schemes, it is convenient to express the normalized response in decibels by using  
$10\log_{10}\left\vert R(\omega)/R(0) \right\vert ^2$, which leads to  
\begin{gather}
\bar{R}_\text{oscillator}\left( f\right)=10\log_{10}\left( |u_3\left(f\right)|^2/|u_3\left(0\right)|^2 \right) ,
\label{eq:R_osc} \\
\bar{R}_\text{laser}\left( f\right) = 10\log _{10}\left( |\delta S_{y0}\left(f\right)|^2 /|\delta S_{y0}\left(f_\text{low}\right)|^2 \right),
\label{eq:R_laser}
\end{gather}
where the frequency is $f = \omega / 2\pi $, while for the laser response
we choose the normalization at a low 
frequency $f_\text{low}=0.43\ \mathrm{GHz}$.
These response functions are 
shown in Fig.~\ref{fig:cho2} with the
corresponding phase shifts 
from the modulation source $\phi_M$ 
as the arguments of complex functions $u_3\propto x_2$ and $\delta S_{y0}$,
given in the two insets. 
Our main goal is to compare the trends in the responses of the two different systems.  
Rather than choosing the parameters
for the best numerical matches of these responses, we consider a simple set of the parameter values describing the masses, spring constants, and damping coefficients for coupled harmonic oscillators and retain them, even when we change the modulation scheme. 

Similarly, the trends in lasers are 
illustrated by keeping a fixed $\gamma_p=75\pi\ \mathrm{GHz}$, a value several times smaller than experimentally realized $\gamma_p>200\pi\ \mathrm{GHz}$~\cite{Lindemann2019:N,Pusch2015:EL}, but large enough
to support dynamical response in spin-lasers which is faster than in their 
best conventional counterparts, recall Eq.~(\ref{eq:fr_tilde}).
Furthermore, this chosen $\gamma_p$ is 
consistent with the grating-induced experimental 
values~\cite{Lindemann2023:EL,Pusch2019:EL}
that offer design flexibility for future scaled-down
lasers with electrical spin injection.
We express the spin relaxation using its rate $\gamma_s=1/\tau_s$.
In all the modulation schemes, we choose $P_{J0}=0.04$. While experiments also support a higher
spin polarization of injected carriers~\cite{Zutic2020:SSC}, this modest value 
already yields a completely circularly polarized light,
for $P_{J0}\approx 0.04$, $P_C >0.95$ at room temperature~\cite{Iba2011:APL}. These results can be understood from the spin-bucket model in Fig.~\ref{fig:cho1}(b): Near the lasing threshold ($J_0 \gtrsim 1$), when the water level is just above the slit, only the hot water (pumped more) will overflow.

By comparing the trends in Figs.~\ref{fig:cho2}(a) and \ref{fig:cho2}(b) we see good agreement in the intensity and phase response of coupled harmonic oscillators and spin-lasers. At low $f$, below the resonances 
in Figs.~\ref{fig:cho2}(a) and \ref{fig:cho2}(b), the amplitudes $x_2$ and $\delta S_y$ oscillate nearly in phase with the modulation source (external force or current) $\phi_M \approx 0$. The maximum energy transfer  
from the modulation source to the oscillating system is at the resonance where 
the phase shift is $\phi_M=\pi/2$. The resonance can be estimated by $\omega_L(k=0)/2\pi \approx 0.5\ \mathrm{Hz}$ from Eq.~(\ref{eq:omegaL}) for the harmonic oscillator and by $\omega_R/2\pi\approx 6.8\ \mathrm{GHz}$ from Eq.~(\ref{eq:omegaR}) for the spin-laser. At the resonance, it can be shown that $\delta S_x$ and $\delta S_y$ as well as $x_1$ and $x_2$ oscillate approximately in phase; their mutual phase difference is zero in the corresponding normal mode $\left( \phi_{M} \right)_x - \left( \phi_{M} \right)_y \approx 0$. At higher $f$, above the resonance, $\phi_M=\pi$ signals that, for both systems, the oscillations are out of phase with the modulation source. With the further increase in $f$, the response functions fall below $-3\ \mathrm{dB}$, beyond the modulation bandwidth.

This could be surprising as, for the considered coupled harmonic oscillators with $k \ll k_2 \ll K_1$, we can obtain the two approximate angular eigenfrequencies 
\begin{eqnarray}
\omega_L(k)\approx\omega_L(k=0)\left(1+\frac{k/2k_2
}{1-c_2^2/\left(4k_2m_2\right)}\right), 
\label{eq:omegaLfull}
\\
\omega_H(k)\approx\omega_H(k=0)\left(  1+\frac{k/2K_1}{1-c_1^2/\left(4K_1 m_1\right)}\right),
\label{eq:omegaHfull}
\end{eqnarray}
but the response is only determined by the lower one 
$\omega_L(k)$, while $\omega_{L,H}(k=0)$ are
defined by Eqs.~(\ref{eq:omegaL}) and (\ref{eq:omegaH}).
Similarly, for the birefringent spin-laser, we only
see the response given by the lower frequency of the
dominant $S_y$ contribution. 
For the considered IM, therefore, it appears that a model 
of a single harmonic oscillator and a single resonance 
is sufficient to describe the trends in the modulation response and the phase shift of a birefringent 
spin-laser. Such a response and the phase shift in the harmonic oscillator are known to arise from the 
interplay between the restoring force, damping, and inertia~\cite{Fowles:2005}.  
It can be shown that lower resonant frequency depends mostly on $1/\tau_\text{ph}$.
The response for a lower frequency determining the dominant $\delta S_y$ oscillations is largely driven 
by $1/\tau_\text{ph}$ which resembles the role of a restoring force. 

\subsection{\label{sec:PM}C. Polarization modulation}

We extend our comparison of the two systems to PM
which is modeled by CASE 2 in Fig.~\ref{fig:cho1}(c) and 
defined for spin-lasers by
\begin{align}
P_J = P_{J0} + \delta P_J\left( t \right), 
\label{eq:PM} 
\end{align}
where we consider harmonic modulation and small-signal analysis 
$\delta P_J \left( t \right) = \operatorname{Re}\left[\delta P_J\left(\omega \right) e^{-i \omega t} \right]$, 
with $\left \vert\delta P_J\left( t \right)\right\vert \ll 1$, 
while $J = J_0$. As before for IM, we can also 
obtain our PM result independent of the magnitude 
of a small modulation ($\delta P_J \ll 1)$. 

Prior PM studies have recognized its various advantages, including an increased bandwidth and ultrafast and energy-efficient operation in 
spin-lasers~\cite{Lee2010:APL,Yokota2018:APL,Lindemann2019:N}.
Specifically, ultrafast
operation in highly birefringent 
spin-lasers can be realized at
low injections $J_T \lesssim J$, 
which has been recently demonstrated
with electrically tunable birefringence, even at elevated 
temperatures
$\sim\!70\ \mathrm{^\circ C}$~\cite{Lindemann2020:AIPA}. This could greatly reduce the power
consumption, which is estimated to be an order of magnitude lower than in the state-of-the-art conventional 
lasers~\cite{Lindemann2019:N}, as well
as mitigate the cooling requirements  
and growing water consumption  
for high-performance interconnects in 
large datacenters~\cite{Mytton2021:npjCW}.
Furthermore, 
PM reduces a parasitic frequency modulation (chirp)
associated with linewidth broadening, enhanced dispersion, and limiting the high bit rate in 
telecommunication systems~\cite{Boeris2012:APL}.

With PM, it is possible to turn the lasing on
and off, even at a fixed $J$. Some of the
trends in the dynamical operation of spin-lasers are different from what is usually expected in spintronics, where $\tau_s/\tau_r \gg 1$ is desirable~\cite{Zutic2004:RMP}. Instead, 
an enhanced bandwidth in PM is consistent with
$\tau_s/\tau_r \ll 1$~\cite{Lee2010:APL,Lindemann2019:N}, while the specific value needs to be optimized, as it comes at the cost of a reduced signal.

Like Eq.~(\ref{eq:response}), the response function for PM is defined by $R(\omega)=|\delta S_y(\omega)/\delta P_J\left(\omega \right)|$. This is different from some
of the other studies, where PM was associated 
with considering the changes in the circular polarization of the emitted light $P_C$, while we focus on the
dynamics of the $y$-polarized dominant mode $S_y$. 
Linearized intensity equations given by Eq.~(\ref{eq:Xlin}), which include a change in $P_J$ from Eq.~(\ref{eq:PM}), take the form
\begin{equation}
\delta\mathbf{\dot{X}} = \bm{\mathsf{M}}\left(  \mathbf{X}_{0},P_{J0}\right)
\delta\mathbf{X}+\mathbf{G}^{\text{PM}}\left(  \mathbf{X}_{0},P_{J0}\right)  \delta P_{J},%
\label{eq:deltaXdot}
\end{equation}
where $\mathbf{G}^{\text{PM}} = \left( G^{\text{PM}}_1, G^{\text{PM}}_2, G^{\text{PM}}_3, G^{\text{PM}}_4 \right)$ and
\begin{eqnarray}
G^{\text{PM}}_1 &=& -\frac{\tau_s}{2\tau_{\text{ph}}}J_0\left[\alpha \sin\phi_0\left(  \sqrt{
S_{y0}/S_{x0}}-\sqrt{S_{x0}/S_{y0}} \right) \right. \nonumber \\
 &&\left.  +\cos\phi_0\left(\sqrt{S_{y0}/S_{x0}}+\sqrt{S_{x0}/S_{y0}}\right)  \right], 
\label{eq:G1_PM} \\
G^{\text{PM}}_2 &=& \frac{\tau_s}{\tau_{\text{ph}}} J_0 \sqrt{S_{x0}S_{y0}}\left(  \alpha \cos\phi_0%
-\sin\phi_0\right), 
\label{eq:G2_PM} \\
G^{\text{PM}}_3 &=& -\frac{\tau_s}{\tau_{\text{ph}}}J_0 \sqrt{S_{x0}S_{y0}}\left(  \alpha \cos\phi_0%
+\sin\phi_0\right), 
\label{eq:G3_PM} \\
G^{\text{PM}}_4 &=& 0,
\label{eq:G4_PM}
\end{eqnarray}
where we see that we recover  
$G^{\text{PM}}_i$, $i=1, 2, 3$, 
by replacing $P_{J0}$ with $J_0$ in 
$G^{\text{IM}}_i$, $i=1, 2, 3$,  in 
Eqs.~(\ref{eq:G1_IM})--(\ref{eq:G3_IM}).
The solution of Eq.~(\ref{eq:deltaXdot}) can be found by using Eq.~(\ref{eq:deltaX}) as
\begin{equation}
\delta\mathbf{X}_0\left(\widetilde{\omega} \right)  = -\left(  \bm{\mathsf{M}}+i \widetilde{\omega}
\bm{\mathsf{I}}\right)  ^{-1}\mathbf{G}^{\text{PM}}\delta P_{J0},
\label{eq:deltaX0omega}
\end{equation}
with the assumption of $\delta P\left(0 \right) = \delta P_{J0}$ being the frequency-independent amplitude.

\begin{figure}[t]
\centering
\includegraphics*[width=8.6cm]{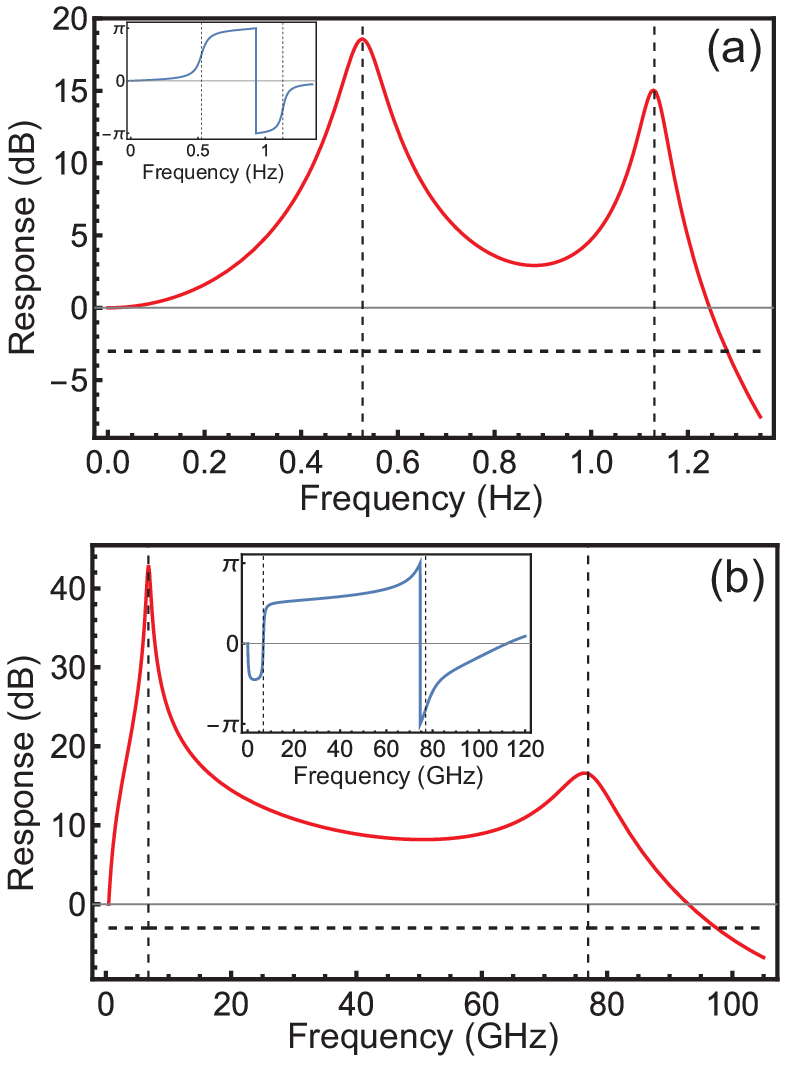}
\vspace{-0.5cm}
\caption{(a) Response function for CASE 2 of the coupled harmonic oscillators from Fig.~\ref{fig:cho1}(c), describing the displacement $x_2$. Dashed horizontal line: 
$-3\ \mathrm{dB}$ modulation bandwidth. 
Vertical lines mark the approximate eigenfrequencies from Eqs.~(\ref{eq:omegaLfull}) and (\ref{eq:omegaHfull}). Inset: Phase evolution. (b) PM response function for $S_y$ and the phase evolution (inset). Vertical lines indicate relaxation oscillation frequencies. The parameters for the coupled harmonic oscillators and spin-lasers are retained from Fig.~\ref{fig:cho2}.
}
\label{fig:cho3}
\vspace{-0.3cm}
\end{figure}

The response functions from Eqs.~(\ref{eq:R_osc}) and (\ref{eq:R_laser}), where $u_3$ is calculated for CASE 2,
while $\delta S_{y0}$ is given in Eq.~(\ref{eq:deltaX0omega}), are shown in Fig.~\ref{fig:cho3}. The plots of the 
corresponding phase shifts are given in the insets. 
In both cases, we see the key difference with IM:
There are two peaks now, and a model of a single harmonic
oscillator is clearly insufficient for PM in spin-lasers.
There are also some differences in the specific 
shapes of the two-peak response functions in Figs.~\ref{fig:cho3}(a) and \ref{fig:cho3}(b), which
also arise from a larger relative frequency separation 
between the two resonances in a spin-laser.
For the considered spring constants, $K_1 \gg k_2 \gg k$ 
($K_1 = 100\,k = 10\,k_2$), the locations of the two peaks in Fig.~3(a) are accurately described by 
$\omega_{L,H}/2\pi$ in Eqs.~(\ref{eq:omegaLfull}) and (\ref{eq:omegaHfull}). This means that, in Fig.~\ref{fig:cho3}(a),
the peak at the lower frequency $\omega_L/2\pi$, 
already recovers the behavior from the single peak 
in CASE 1, while the new peak appears at $\omega_H/2\pi$.

There is a similar situation with the PM, where we 
can also identify the origin of the two peaks. The one
at a lower frequency can again be estimated as in the
IM case with $\omega_R/2\pi$ from Eq.~(\ref{eq:omegaR}). 
It corresponds
to the dominant $y$ mode and is largely independent 
of $\gamma_p$. In contrast, the peak at the higher frequency $f_R^\text{PO}$ from Eq.~(\ref{eq:fr_tilde}) is driven by $\gamma_p$. While the coupling between $\delta S_x$ and $\delta S_y$ is weak, it is still sufficient to transfer a part of the oscillation energy from $\delta S_x$ to $\delta S_y$ and to excite the normal mode which corresponds to $f_R^\text{PO}$.
In prior PM experiments using
optically injected spin,
$f_R^\text{PO} \approx 27 f_R$ with $\gamma_p\approx 212 \pi\ \mathrm{GHz}$~\cite{Lindemann2019:N}. Our results confirm that, 
in highly birefringent lasers, the PM
supports not only ultrafast dynamics in $P_C$ but also in
the dominant lasing mode. 

If we compare the phase evolution in the insets of 
Figs.~\ref{fig:cho3}(a) and \ref{fig:cho3}(b), we find they both display multiple
phase jumps, but there are also noticeable differences.
For coupled harmonic oscillators, there are two
$\pi$ jumps at $\omega_L(k)/2\pi$ and $\omega_H(k)/2\pi$,
while for a spin-laser, we see a more complicated 
behavior which also includes a low-frequency dip.
Starting at $f=0$, the phase shift abruptly falls to  $\phi_M \approx -\pi/2$ before the lower resonance and, therefore,
$\delta S_y$ leads $\delta P_J$. The explanation for such behavior, as we show in Sec.~IV, can be found by considering variables $\left(\phi,S_x\right)$ and $\left(S_y,N\right)$ as two independent oscillators in the leading 
approximation. The dynamics of $\delta S_y$ for 
$f \ll f^\text{PO}_R$ ($f\lesssim 10\ \mathrm{GHz}$) is well described by
\begin{eqnarray}
&\delta\ddot{S}_y& - \left(  M_{33}+M_{44}\right)  \delta\dot{S}_y+\left(
M_{33}M_{44}-M_{34}M_{43}\right)  \delta S_y \nonumber \\
&&=\left(  M_{34}G_4^{\text{PM}}
-M_{44}G_3^{PM}\right)  \delta P_J+G_3^{\text{PM}}\delta\dot{P}_J,
\label{eq:deltaSydots}
\end{eqnarray}
where both $M_{ij}$, the elements of matrix 
$\bm{\mathsf{M}}$, as well as the components $G_3^{\text{PM}}$ and $G_4^{\text{PM}}$ are real. For
harmonic PM,  
$\delta \dot{P}_J = -i \widetilde{\omega}\delta P_J$ in Eq.~(\ref{eq:deltaSydots}).  At low $f \approx 1\ \mathrm{GHz}$, the imaginary part is of the same order of magnitude as the real part on the right-hand side of Eq.~(\ref{eq:deltaSydots}) for PM, producing the phase shift for $\delta P_J$. However, 
for IM or coupled harmonic oscillators in CASE 1 and CASE 2, the modulation source acquires no additional phase shift.

After $f$ passes
through the lower resonance with $\phi_M \approx 0$, the phase shift 
 rises to its maximum value of $\pi$. With a further increase in $f$, $\phi_M$ drops to $-\pi$ and then, at a higher resonance $f = f^\text{PO}_R$, $\phi_M \approx -3\pi/4$, which again is because $\delta P_J$ picks up an extra phase shift. In contrast with CASE 1 and IM, after the resonance at the higher frequency, both $x_2$ and $\delta S_y$ are almost in phase with the modulation source for 
$f \approx f_\text{3dB}$. When $f \rightarrow \infty$, for $\delta S_y$,  there is $\phi_M \rightarrow  \pi/2$.  
Due to the geometric arrangement of masses and external force, it can be shown that $x_1$ lags $x_2$ in CASE 1 and leads it in CASE 2 at $\omega_H(k)/2\pi$. This is unlike in a spin-laser, for which $S_x$ always leads $S_y$ at $f^\text{PO}_R$.

\subsection{\label{sec:IM+PM}D. Intensity modulation and polarization modulation}

The previous examples of IM and PM are just two of 
many possible modulations in spin-lasers.
In fact, for specialized applications, some other 
modulation schemes could be more advantageous~\cite{Boeris2012:APL}. It is
then tempting to also simultaneously consider IM and PM for spin-lasers.
\begin{figure}[t]
\centering
\includegraphics*[width=8.6cm]{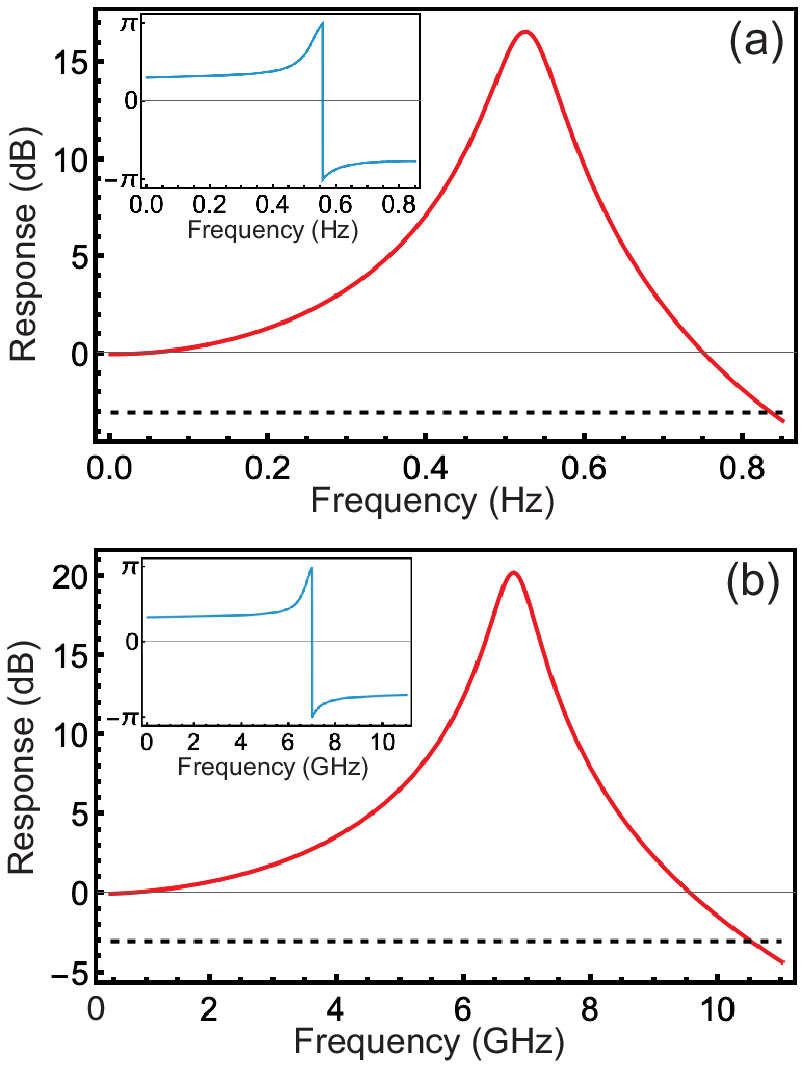}
\vspace{-0.5cm}
\caption{(a) Response function for $x_2$ with a combined modulation of the coupled harmonic oscillators from 
Fig.~\ref{fig:cho1}(c). For CASE 2, the fixed point $P$
is replaced by a harmonic force of a different amplitude and phase than at the point $O$, described by 
$\mu =\exp \left( i\pi/3 \right )$.
(b) Response function for $S_y$ with a simultaneous
IM and PM for a spin-laser. The relative modulation contributions are described by 
$\xi = \exp \left( i\pi/3 \right )$.
The remaining parameters in (a) and (b)
are taken from Fig.~\ref{fig:cho2}.
}
\label{fig:cho4}
\vspace{-0.3cm}
\end{figure}
To examine that, 
and if it can offer
further improvements in spin-lasers, it would help
to again derive some intuition from the model of
coupled harmonic oscillators. In the scheme from Fig.~\ref{fig:cho1}(c),
such a simultaneous modulation implies that we need
to simultaneously consider harmonic forces at both ends:
at points $O$ and $P$, by removing the left (right) fixed
point in CASE 1 (CASE 2). Using our previous
approach and combining CASE 1 and CASE 2, for the 
coupled harmonic oscillators we can obtain the
solutions for amplitudes of the displacement 
\begin{equation}
\mathbf{u}\left(  \omega' \right) = -\left( \bm{\mathsf{U}} + i\omega' \bm{\mathsf{I}}
\right)^{-1} \left( \mathbf{b}_2 + \mu \mathbf{b}_1 \right),
\label{eq:uomega'}
\end{equation}
where $\mu =  B/B'$ 
is the ratio between the amplitudes of the displacement due to external forces in CASE 1 and CASE 2.
With a similar approach of a
combined IM and PM applied to spin-lasers, 
we obtain
\begin{equation}
\delta\mathbf{X}_0\left(\widetilde{\omega} \right)  =  -\left(  \bm{\mathsf{M}}+i \widetilde{\omega}
\bm{\mathsf{I}}\right)^{-1}\left(\mathbf{G}^{\text{PM}} + \xi \mathbf{G}^{\text{IM}} \right)\delta P_{J0},
\label{eq:deltaX0omegaT}
\end{equation}
where $\xi = \delta J_0/\delta P_{J0}$, as
with $\mu$ for harmonic oscillators, describes the
dominance of one modulation scheme. Both
$\mu$ and $\xi$ are complex and include the phase difference between the two sources of a harmonic modulation. 

The plots of response functions for $u_3$ and $\delta S_{y0}$ in Eqs.~(\ref{eq:uomega'}) and (\ref{eq:deltaX0omegaT}) are shown in Fig.~\ref{fig:cho4} for equal amplitudes of the two sources $|\mu|=1$ and equal amplitudes of IM and PM $|\xi|=1$.
Even with the additional contributions
from CASE 2 and PM, the corresponding response functions show only a single peak in the displacement and photon density with the same shape as obtained for CASE 1 and IM in Fig.~\ref{fig:cho1}. However, the phase shift 
information is now different. For both coupled harmonic
oscillators and spin-lasers, at $f=0$, we see that $\phi_M=\pi/3$ since there is a phase difference between the two modulation sources. 
This phase difference is also the cause for a sudden jump near the resonance because $\phi_M$ cannot exceed $\pi$.

Surprisingly, the response function in Fig.~\ref{fig:cho4}(b) shows no visible change from IM in Fig.~\ref{fig:cho2}(b), even if a different phase is chosen for $\xi$ or,  if $\left\vert \xi \right\vert = 1/100$, the magnitude of the $J$ modulation is only 1\% of the $P_J$ modulation. 
It appears that the combined  
IM and PM
is not as effective as PM to achieve an enhanced bandwidth; a higher-resonance frequency cannot be reached. This disappearance of a higher-resonance peak can be understood as follows: Oscillations
in coupled harmonic oscillators are strongly affected by inertia and restoring force at higher frequencies. Mass $m_2$ oscillates as a free particle in the presence of two external forces. In addition to the force connected to point $P$, now $m_1$ pulls and pushes on $m_2$, contributing significantly only near the resonance because of the much smaller displacement amplitude. 

Away from the resonance, the force on the right side of $m_2$ dominates, and  
$u_3$ decreases as $1/\omega^2$, just like for CASE 1. Analogously, at higher $f$,  $\delta S_x$ and $\delta J$ are
acting upon $\delta S_y$, where $\delta S_x$ is 
pushed and pulled by $\delta P_J$ and  the
effective restoring force is controlled by $\gamma_p$. However, $\delta J$ generates a much larger response $\delta S_{y0}$ than the other $\delta P_J$
modulation, producing almost the same response as for 
IM alone. While the enhanced bandwidth is not feasible
with the considered IM and PM, we see that
such a modulation scheme could be used to tailor the phase evolution and enable other opportunities for
the transfer and processing of information~\cite{Csaba2020:APR}.
Nevertheless, 
for a much larger range of the ratios for the IM and PM amplitudes, we find in the Appendix that the second peak and an enhanced $f_\text{3dB}$,  
missing in Fig.~\ref{fig:cho4}, are both restored.  
The trends found from the coupled harmonic oscillators provide 
the guidance for spin-lasers.

A long history of using oscillator-based computing and the crucial role of phase information can already be
seen from the realization of a parametron~\cite{Goto1959:PIRE}, which has not only preceded integrated circuits but continues to inspire various von Neumann computing architectures~\cite{Csaba2020:APR}.

\section{\label{sec:Qfac}IV. Quality and Coupling Factors}

The energy loss in damped harmonic oscillators or resonators is commonly described by their quality factor $Q$~\cite{Fowles:2005}.
It is defined as the ratio of 2$\pi$ times the 
energy stored in an oscillator or resonator and the
energy lost in a single period of oscillations. Here $Q$ is
also used to characterize the resonance shape; its full width
at half-maximum $\Delta \omega$ is $\omega_0/Q$ or, equivalently~\cite{Jackson:1999},
\begin{equation}
Q=\omega_0/\Delta \omega,
\label{eq:Q}
\end{equation}
where $\omega_0$ is
the angular resonance frequency without losses. Different parts of a laser, for example, its resonant cavity and the gain region, have different $Q$ factors. 

\begin{figure}[t]
\centering
\includegraphics*[width=8.6cm]{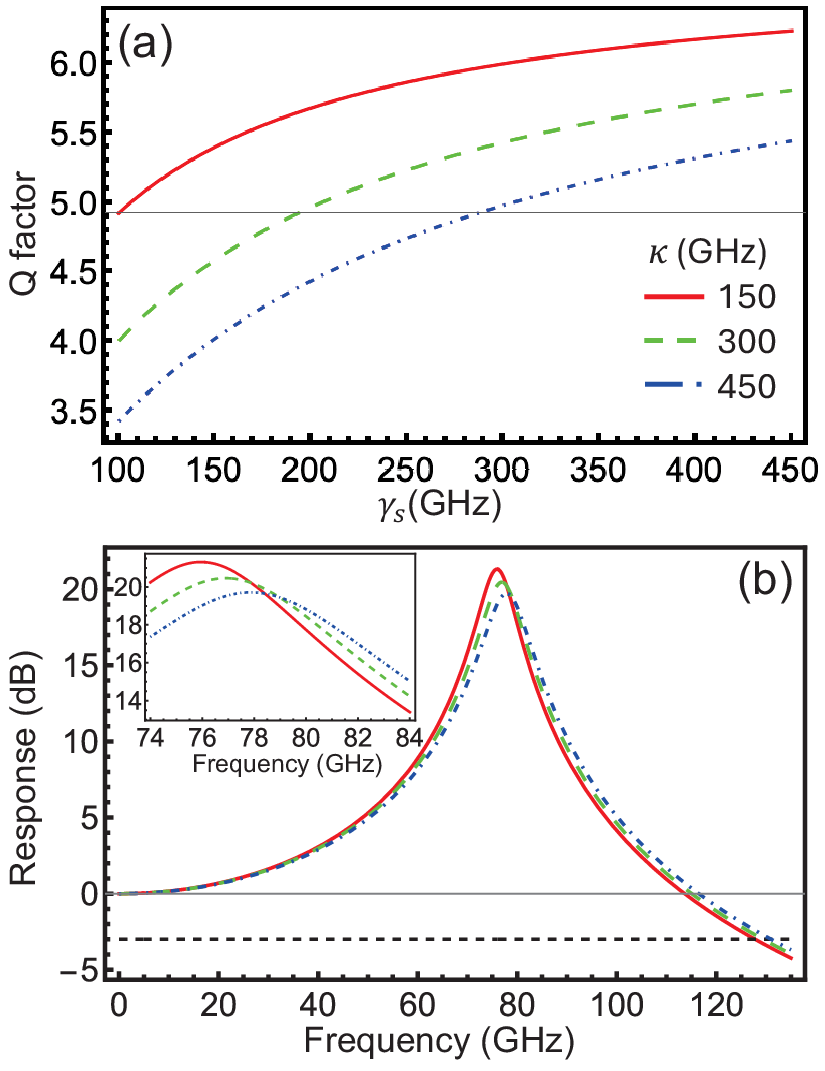}
\vspace{-0.5cm}
\caption{(a) $Q$ factor for PM response of the emitted light polarization $P_C$ as a function of the spin-relaxation rate $\gamma_{s}$ 
for three 
values of photon decay rate, $\kappa=1/2\tau_\text{ph}$. 
(b) PM response for $P_C$ 
for the values of $\kappa$ in (a). 
Inset: Magnified region near the peak response.
The other spin-laser parameters are taken from Fig.~\ref{fig:cho2}.
}
\label{fig:cho5}
\vspace{-0.3cm}
\end{figure}

We focus on the $Q$ factor for the gain region and the polarization of the emitted light $P_C$ from Eq.~(\ref{eq:PC}) at the higher-resonant frequency. The coupled harmonic oscillator model for a spin-laser can help us to elucidate the role of key parameters in the intensity equations.
Photon densities for the two  
helicities can be expressed
in terms of linear polarizations and their relative phase as
\begin{equation}
S^{\pm } = \left( S_x+S_y\pm \sqrt{S_xS_y}\sin \phi
\right)/2.
\label{eq:Spm}
\end{equation}
Within the small-signal analysis, for PM we have 
\begin{equation}
\delta P_{C}=2\left(S_0^-\delta S^+-S_0^+\delta S^-\right)/\left( S_0^++S_0^-\right)^2,
\label{eq:dPC}
\end{equation}
where $S_0^+$ and $S_0^- $ are the steady-state values 
and, as before, we assume the harmonic change
in time for $\delta S^-$, $\delta S^+$, and $\delta P_C$ in Eq.~(\ref{eq:dPC}). 
The steady-state value $P_{C0} \approx 0$ and the polarization of light is almost linear. 
\begin{figure}[ht]
\centering
\includegraphics*[width=8.6cm]{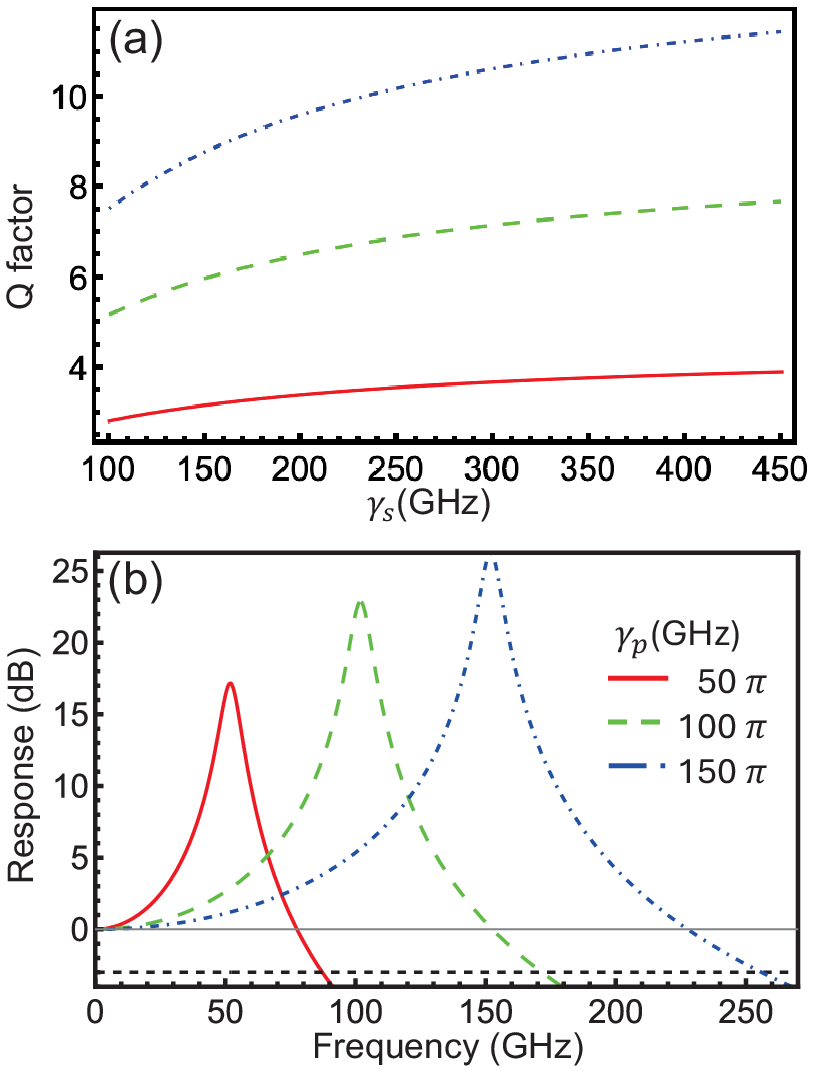}
\vspace{-0.5cm}
\caption{(a) $Q$ factor for PM response for $P_C$ as a function of spin-relaxation decay rate for three 
values of birefringence $\gamma_{p}$. (b) PM response function $P_C$ for three values of $\gamma_{p}$. 
The other spin-laser parameters are taken from Fig.~\ref{fig:cho2}.
}
\label{fig:cho6}
\vspace{-0.3cm}
\end{figure}

By calculating $Q$ from Eq.~({\ref{eq:Q}) and the PM response 
for $P_C$:
\begin{equation}
R(\omega) = |\delta P_C(\omega)/\delta P_J (\omega)|,
\label{eq:PMPC}
\end{equation}
normalized to its zero-frequency value as in Eq.~(\ref{eq:R_laser}), 
we examine their different evolution with 
$\gamma_p$, the spin-relaxation rate $\gamma_s$, and 
the photon decay rate $\kappa = 1/2\tau_\text{ph}$, each of them changing threefold. 
In Fig.~\ref{fig:cho5}(a), we see an expected change for the 
increase of $\kappa$ on the evolution of $Q$ with $\gamma_s$: A 
smaller $\kappa$ is responsible
for an increase in $Q$, just as if the losses are reduced. At each $\gamma_s$, this increase is 
sublinear $\lesssim 50$\%
for a threefold change in 
$\kappa.$  In contrast, $Q$ increases (sublinearly) with $\gamma_s$,
so the spin-relaxation decay rate is not
a simple loss mechanism. From Fig.~\ref{fig:cho5}(b) and its inset, we see that $\kappa$ has
only a modest influence on the $P_C$ response function and slightly increases the frequency for the peak response. While the
PM response for $P_C$ in Fig.~\ref{fig:cho5}(b) with $\kappa=300\;$GHz
employs the same laser parameters as for the the PM response for $S_y$ in Fig.~\ref{fig:cho3}(b),
there are striking differences. The $P_C$ response has only a single peak at the higher frequency, given by
Eq.~(\ref{eq:fr_tilde}), while the lower peak 
from the $S_y$ response is completely absent. 
This behavior could be inferred from the phase evolution of the analogous model of coupled harmonic oscillators.
A more detailed explanation can
be obtained from the analysis of Eq.~(\ref{eq:dPC}).
At $J_0=4J_T$, the  
analogy
from Fig.~\ref{fig:cho1}(b) suggests that 
there is an overflow of both hot and cold water: The lasing is realized with both helicities of light and $P_{C0} \approx 0$~\cite{Gothgen2008:APL}. The emitted light is 
almost linearly polarized $S_0^+\approx S_0^-$. At the lower resonance, one can also show that $(\delta S^+ - \delta S^- )\sim \delta S_x$
has only one peak at the higher resonance, 
the same behavior as the response of $x_1$ for CASE 2. 
From Eq.~(\ref{eq:dPC}), we can conclude
that $\delta P_C$ is negligible, and there is no lower-resonance peak.

Instead of changing $\kappa$, in Figs.~\ref{fig:cho6}(a) and ~\ref{fig:cho6}(b), we examine the same evolution of $Q$ and  the $P_C$ response function for different $\gamma_p$. Both of these quantities clearly increase with $\gamma_p$, which modifies them more ($\sim$5 times) than with three fold changes in $\gamma_s$ or $\kappa$. A similar trend for the single-peak $P_C$ response function with $\gamma_p$ is obtained from a generalized SFM at $J_0=1.5J_T$, even up to $\gamma_p=250\pi$~\cite{Lindemann2019:N}. The validity of our small-signal analysis from the intensity-rate equations in Sec.~IIIC is further verified from the experimental results for spin-lasers where the increase of $\gamma_p=7, \,15.7,\, 21.1\;$GHz was  realized by their mechanical deformation and accompanied by an increase in the resonance frequency. These moderate $\gamma_p$ values were chosen to allow for the detection of the modulation response~\cite{Lindemann2019:N}.

Taken together, for the considered parameter range, we can conclude that $\gamma_p$ shows a similar behavior to the restoring force, while $\kappa$ has mostly the character of the damping constant. Here, $Q$ decreases with $\kappa$, while we see a slight increase in $f_\text{3dB}$ with $\kappa$ so the damping associated with $\kappa$ is not completely detrimental. In contrast, both $Q$ and $f_\text{3dB}$ clearly increase with $\gamma_s$, which suggests that $\gamma_s$ has the role of the restoring force or even inertia, within the analogy with the model of coupled harmonic oscillators. We will return to this connection as we next quantify the coupling strength of the oscillators.

A closer analysis of our model of spin-lasers as coupled harmonic oscillators offers further insights that they are weakly coupled.
Matrix $\bm{\mathsf{M}}$ can be decomposed in a similar way as the matrix
$\bm{\mathsf{U}}$ in Sec.~II
\begin{equation}
\bm{\mathsf{M}} = \bm{\mathsf{T}}_{\mathsf{0}} + \bm{\mathsf{W}},
\label{eq:MTW}
\end{equation}
where $\bm{\mathsf{T}}_{\mathsf{0}}$ ($\bm{\mathsf{W}}$) is a block diagonal
(block off-diagonal) matrix. The blocks are $2\times 2$ matrices but are 
considerably more complicated than the blocks displayed in Eqs.~(\ref{eq:Hmatrix}) and (\ref{eq:Vmatrix}). 
Matrix $\bm{\mathsf{T}}_{\mathsf{0}}$ describes approximately two independent (decoupled) oscillators: (i) corresponding to  variables $\left( \phi ,S_x\right)$, which oscillates with
a higher frequency $\Omega _H/2\pi$, and the damping parameter $\Gamma _H$, where
\begin{align}
\Omega _H^2& =M_{11}M_{22}-M_{12}M_{21} ,
\label{eq:OmegaH2}
\\
\Gamma _H& =-\left( M_{11}+M_{22}\right),
\label{eq:GammaH}
\end{align}
and (ii)  described by $\left( S_y,N\right)$, oscillating with a lower frequency $\Omega _L/2\pi$ and the damping parameter $\Gamma _L$, where
\begin{align}
\Omega _L^2& =M_{33}M_{44}-M_{34}M_{43}, 
\label{eq:OmegaL2}
\\
\Gamma _L& =-\left( M_{33}+M_{33}\right),
\label{eq:GammaL}
\end{align}
and $M_{ij}$ is calculated from Eqs.~(\ref{eq:F1})--(\ref{eq:F4}) and (\ref{eq:Mij}).

\begin{figure}[t]
\centering
\includegraphics*[width=8.6cm]{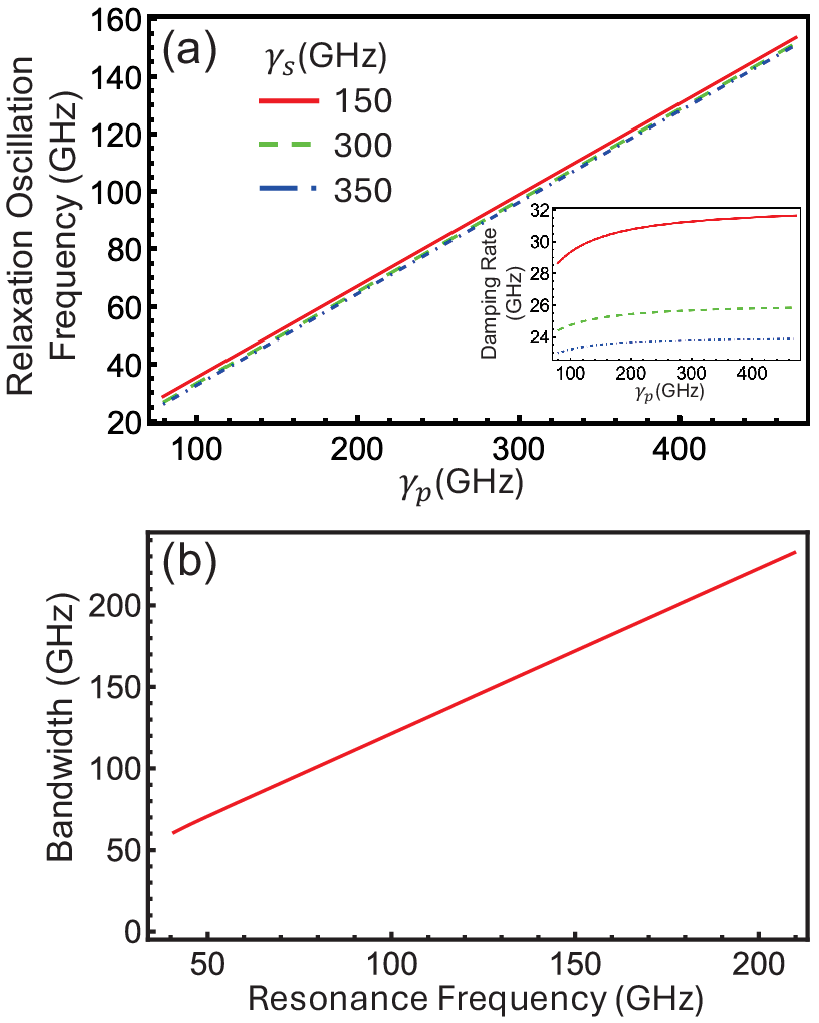}
\vspace{-0.5cm}
\caption{(a) Higher relaxation oscillation frequency $f_H$ 
and the corresponding damping rate $\gamma_H$ (inset) as a function $\gamma_p$ for different $\gamma_s$. (b) PM bandwidth $f_{3\text{dB}}$
calculated for the response function $\delta S_{y0}\left( f\right)$
is almost linear in the higher resonance frequency $f_R^\text{PO}$ [see Eq.~(\ref{eq:fr_tilde})],
$f_{3\text{dB}} \approx 20\ \mathrm{GHz} + f_R^\text{PO}$.
The other spin-laser parameters are taken from Fig.~\ref{fig:cho2}.
}
\label{fig:cho7}
\vspace{-0.3cm}
\end{figure}

Eigenvalues of 
$\bm{\mathsf{M}}$ provide information about the relaxation oscillations in spin-lasers, important for signal modulation and stability. 
The positive imaginary part of the eigenvalues corresponds to the angular frequency, while the real part gives the negative damping rate of the relaxation oscillations $\gamma$. While their long expressions 
can be calculated analytically,  
it is useful to consider a more transparent, approximate form 
using a series expansion for eigenfrequencies of $\bm{\mathsf{T}}_0$ in Eq.~(\ref{eq:MTW}) for $S_x \ll 1$
\begin{eqnarray}
\omega _L^2 &\approx &-(1/4)\left[1+2\gamma_a+\left(N_0-1\right)/\tau_\text{ph}\right]^2 \nonumber \\
&& + S_{y0}^2\left(-1/4+\epsilon_{yy}+\epsilon_{yy}^2\right) \nonumber \\
&& + S_{y0}\left[-1/2+\epsilon_{yy}+\gamma_a\left(2\epsilon_{yy}-1\right) \right. 
\nonumber \\
&& \left. 
+\left(N_0+1\right)/2\tau_\text{ph} +\epsilon _{yy}
\left(N_0-1\right)/\tau_\text{ph} \right], 
\label{eq:omegaL2} \\
\omega _H^2 &\approx &\left(J_{0}P_{J0}\tau_s/2\tau_\text{ph}%
\right) ^2\left[ 2\alpha \sin \left(2\phi_0\right) \right. \nonumber \\
&& \left. -\left(\alpha^2-1\right) \cos \left( 2\phi_0\right) \right] S_{y0}/S_{x0},
\label{eq:omegaH2}
\end{eqnarray}
where $\omega_L$ ($\omega_H$) is the lower (higher) relaxation oscillation
angular frequency that depends on all parameters implicitly through 
the steady-state solution
$\left(\phi_0, S_{x0}, S_{y0}, N_0 \right)$.

For a higher frequency $f_H=\omega_H/2\pi$ and the corresponding damping rate $\gamma_H\approx \Gamma_H/2$, our results are shown in Fig.~\ref{fig:cho7}. We see in Fig.~\ref{fig:cho7}(a) a 
linear increase of $f_H \approx f_R^\text{PO}$ with $\gamma_p$, largely
independent for the considered $\gamma_s$. In the inset, the damping rate $\gamma_H$ 
only weakly increases with $\gamma_p$ and weakly decays with $\gamma_s$. In 
contrast, the lower counterpart (not shown) $f_L= \omega_L/2\pi$, which we 
associate with IM, is unaffected by $\gamma_p$. From Fig.~\ref{fig:cho7}(b), we 
see that not only $f_H \propto \gamma_p$ but also the modulation bandwidth 
$f_\text{3dB} \propto \gamma_p$.
However, as we found from a generalized 
SFM~\cite{Lindemann2019:N}, for 
$\gamma_s \ge 2\gamma_p/\pi$, an approximate linear increase in
$f_\text{3dB}$ with $f_R$ is preserved, the modulation
response for $f < f_R$ remains above $-3\ \mathrm{dB}$. As $\gamma_p$ is further increased such that $f_R \lesssim 210\ \mathrm{GHz}$, for the used 
values of $\gamma_s$, there is a region at $f<f_R$ where the 
response drops below $-3\ \mathrm{dB}$, and $f_\text{3dB}$ no longer increases with $\gamma_p$. To restore a further desirable increase of $f_\text{3dB}$ with $\gamma_p$, one should seek gain materials with a larger 
$\gamma_s$~\cite{Lindemann2019:N}.

If we substitute the standard values of the parameters in the expressions for a spin-laser, we can confirm that the values of frequencies in Eqs.~(\ref{eq:OmegaH2}) and (\ref{eq:OmegaL2}), 
eigenfrequencies of $\bm{\mathsf{T}}_0$ and resonant frequencies and eigenfrequencies of $\bm{\mathsf{M}}$, that is, relaxation oscillation frequencies, differ within $5 \%$ for 
a standard range of the considered parameter values. Such a result suggests that the perturbation $\bm{\mathsf{W}}$ in Eq.~(\ref{eq:MTW}) 
is small, the oscillators are weakly coupled and the damping is small. Nevertheless, it is crucial to recognize that the dynamic equation for $S_y$ as an independent oscillator, given by  Eq.~(\ref{eq:deltaSydots}), 
is a good approximation only in the low-frequency regime, where the lower-resonance frequency is found.

What is considered a weak or a strong coupling in any system of coupled harmonic oscillators, both in classical and quantum regimes, can be identified by two mutually
exclusive criteria. Therefore, a different expression should be used to estimate the magnitude of the coupling~\cite{Rodriguez2016:EJP}. 
We employ the following condition: If the energy exchange rate exceeds the difference between the loss rate of oscillators, then the coupling is strong.
Otherwise, we can consider the system  to be weakly coupled. 
Consequently, the coupling factor is defined as
\begin{equation}
C = \left\vert 2g /\Delta \right\vert,
\label{eq:C}
\end{equation}
where we estimate $g\simeq -\max \{ \vert \Omega_{W} \vert^{2}\} /2\bar{\omega}$. These quantities $\Delta $ and $\bar{\omega}$ are the difference (detuning) 
and the average value of the complex eigenvalues of individual oscillators, respectively, and $\max \{ \vert \Omega_{W} \vert^{2} \}$ is the largest square of 
the absolute value of eigenfrequencies for  $\bm{\mathsf{W}}$. If $C \ll 1$, then the coupling is weak; if $C \gtrsim 1$, then the system is strongly coupled. 
According to this definition, while VCSELs are typically operating in a weakly coupled regime, an important counterexample is their polarization switching~\cite{Michalzik:2013},
which represents a strong coupling.

\begin{figure}[t]
\centering
\includegraphics*[width=8.6cm]{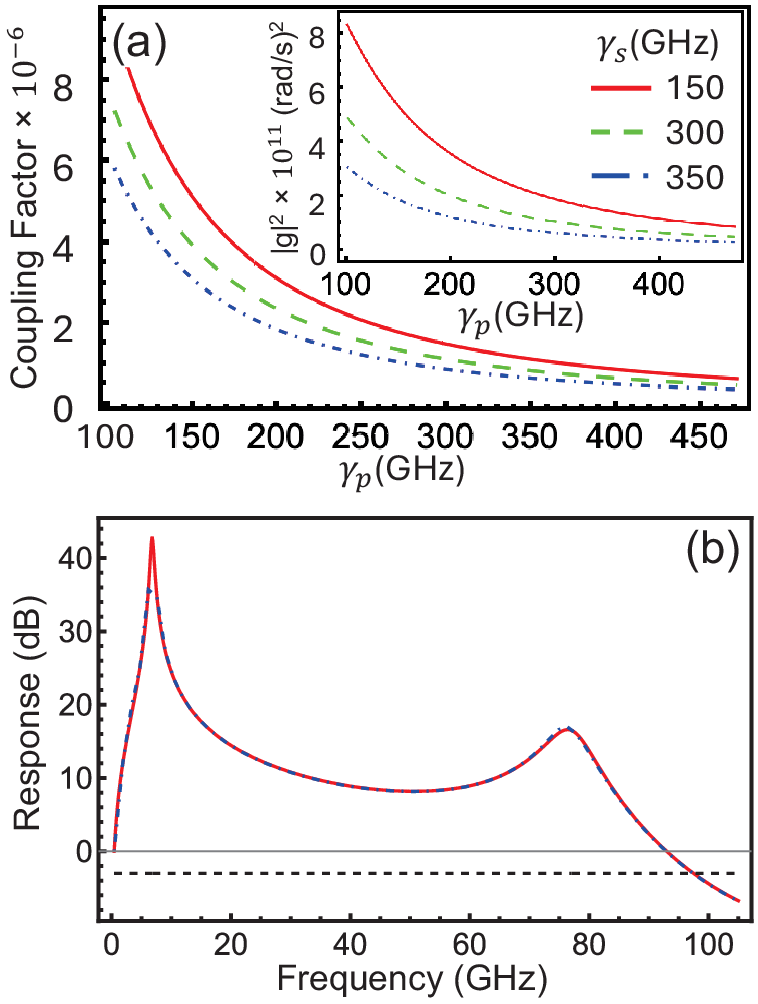}
\vspace{-0.5cm}
\caption{(a) Coupling factor as a function of $\gamma_{p}$ for several values of $\gamma_{s}$. (b) PM
response function for $S_{y}$, numerically calculated from the intensity equations
(dash-dotted line) is compared with the PM response of $\delta S_{y0}$ in Eq.~(\ref{eq:deltaX0omega}), where the term on the left-hand
side is approximated by the first two terms on the right side of Eq.~(\ref{eq:TIW}) (solid line). 
The other lasers parameters are taken from Fig.~\ref{fig:cho2}.
}
\label{fig:cho8}
\vspace{-0.3cm}
\end{figure}

We plot the coupling factor as a function of $\gamma_p$ for several values of $\gamma_s$ in Fig.~\ref{fig:cho8}(a). The approximate complex eigenvalues of the individual oscillators
$-\Gamma_H /2 \pm i\Omega_H$ and $-\Gamma_L /2 \pm i\Omega_{L}$ are calculated using 
Eqs.~(\ref{eq:OmegaH2})--(\ref{eq:GammaL}). For comparison, 
an estimated coupling factor for coupled harmonic oscillators in Fig.~\ref{fig:cho1}(c) is $C \simeq 0.05$, several orders of magnitude larger than for a spin-laser.

We  can also estimate an interaction between independent oscillators 
by relating the middle spring constant $k$ to $\gamma_{p}$. A larger $k$
gives a stronger interaction in the coupled harmonic oscillators system, but a larger $\gamma_p$ results
in the weaker interaction in spin-lasers.  
By substituting%
\begin{equation}
k/m\simeq\left\vert g\right\vert ^2,
\label{eq:k/m}
\end{equation}
we obtain the plot shown in the inset of Fig.~\ref{fig:cho8}(a). As predicted, $k$
decreases with $\gamma_p$. Large values for $\vert g \vert ^2$ are due to $f$ in the gigahertz range.

The perturbative expansion for amplitudes relies on the condition $C \ll 1$ for weakly coupled oscillators. By regarding $\bm{\mathsf{W}}$ as a 
perturbation we expand the matrix $\left( \bm{\mathsf{M}}+i \widetilde{\omega} \bm{\mathsf{I}}%
\right) ^{-1}$ in the Eqs.~(\ref{eq:deltaX0}), (\ref{eq:deltaX0omega}) and (\ref{eq:deltaX0omegaT}) as
\begin{align}
\ & \left( \bm{\mathsf{T}}_{\mathsf{0}} + i \widetilde{\omega} \bm{\mathsf{I}}+\bm{\mathsf{W}}\right)
^{-1} = \left( \bm{\mathsf{T}}_{\mathsf{0}} + i \widetilde{\omega} \bm{\mathsf{I}}\right) ^{-1}
\nonumber \\
\ & -\left( \bm{\mathsf{T}}_{\mathsf{0}} + i \widetilde{\omega} \bm{\mathsf{I}}\right) ^{-1}%
\bm{\mathsf{W}}\left( \bm{\mathsf{T}}_{\mathsf{0}} + i\widetilde{\omega} \bm{\mathsf{I}}\right)
^{-1} \label{eq:TIW} \\
\ &  +\left( \bm{\mathsf{T}}_{\mathsf{0}} + i\widetilde{\omega} \bm{\mathsf{I}}\right) ^{-1}%
\bm{\mathsf{W}}\left( \bm{\mathsf{T}}_{\mathsf{0}} + i\widetilde{\omega} \bm{\mathsf{I}}\right)
^{-1}\bm{\mathsf{W}}\left( \bm{\mathsf{T}}_{\mathsf{0}} + i\widetilde{\omega} \bm{\mathsf{I}}%
\right) ^{-1}+.... \nonumber
\end{align}
The solutions for the amplitudes $\delta\mathbf{X}_0\left(\widetilde{\omega}\right)$ are simplified by keeping only the first two terms on the right-hand
side 
in Eq.~(\ref{eq:TIW}). We note that the inverse of a block diagonal matrix is also a block diagonal.  
We plot this approximate solution for $\delta S_{y0}$ 
and compare it with the small-signal analysis of the PM from the numerical solution of the exact intensity equations in Fig.~\ref{fig:cho8}(b). 
The coupling factor $C \ll 1$ and the approximation of a  small $\bm{\mathsf{W}}$ 
is accurate; the two oscillators $\left(\phi,S_x\right)$ and 
$\left(S_y,N\right)$ are weakly coupled in the intensity equations describing 
a spin-laser. 
The agreement between the two curves in Fig.~\ref{fig:cho8}(b) is good, which
confirms again that the coupling in the system is weak.
This agreement becomes better (worse) for a larger (smaller) $\gamma_p$. 

\section{\label{sec:conc}V. Conclusions and Outlook}

Within this work, we have shown that, by establishing two 
models of asymmetric coupled harmonic oscillators, we can accurately recover the main trends in the IM and PM
of spin-lasers. To recognize the implications of these emerging lasers, we recall that an explosive growth
of artificial intelligence, high-performance computing, and big data reveals that the
advances and power consumption are increasingly determined not only by scaled-down transistors and information processing but also by the energy used in interconnects for information 
transfer~\cite{Wang2023:N,Miller2017:JLT,Jones2018:N}, where spin-lasers can play an important role~\cite{Lindemann2019:N,Dainone2024:N}.

The intuition and analytical results available for the coupled harmonic oscillators 
offer insights not only to better understand the operation of existing spin-lasers but also to elucidate their unexplored regimes. 
There is close agreement between the trends obtained for IM and PM for spin-lasers and the guidance provided by CASE 1 and CASE 2, including the origin of the two resonant frequencies. 
While the single peak in the resonant behavior is a hallmark of IM and CASE 1, for PM, we can understand why it can provide a different behavior at a higher frequency, a double or single peak, depending on if the response is considered for $S_y$ or $P_C$. The analogy with the coupled harmonic oscillators allows us to better understand the role of different material parameters on the quality factor of the considered spin-lasers.

Coupled harmonic oscillators also offer guidance
for unexplored regimes in spin-lasers.
The combination of IM and PM allows exploring not only the transition from a single- to double-peak response, but tailoring the phase evolution, which can be significantly different from the  IM and PM cases alone, even the amplitude of the response is largely unchanged. The importance
of such mixed modulation schemes was suggested
for the elimination of chirp, a parasitic frequency change~\cite{Boeris2012:APL}.

\begin{figure}[ht]
\centering
\includegraphics*[width=8.6cm]{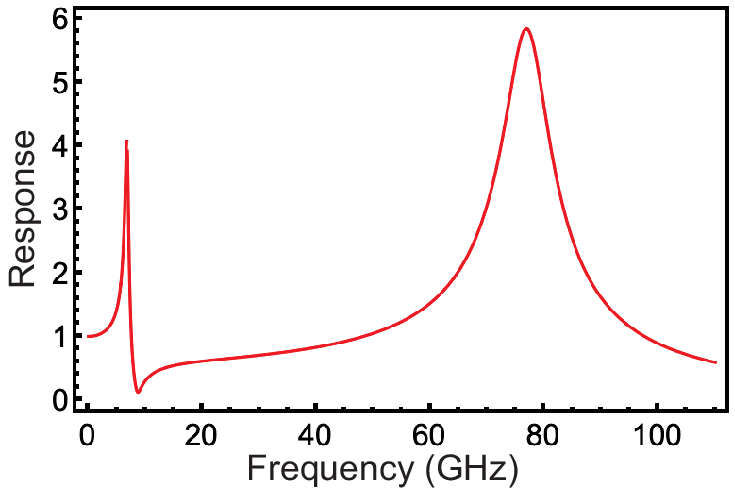}
\vspace{-0.5cm}
\caption{IM response function $\left\vert \delta S_{x0}\left( f\right) /\delta S_{x0}\left(0 \right) \right\vert$ for a spin-laser. Near the lower resonance frequency, at $6.8\ \mathrm{GHz}$, the curve has the characteristic asymmetrical shape of Fano resonance caused by antiresonance at $8.8\ \mathrm{GHz}$. The other spin-laser parameters are taken from Fig.~\ref{fig:cho2}. 
}
\label{fig:cho9}
\vspace{-0.3cm}
\end{figure}
By recognizing the geometry of the coupled oscillators from Fig.~\ref{fig:cho1}(c), we can understand that the spatial position of the driving force, being closer or farther from the given mass will also alter its dynamical evolution. Seemingly, we could then expect that CASE 1, where mass $m_1$ is positioned away from the driving force, would have similarity to CASE 2 and the evolution of $x_2$, describing the position of $m_2$ which, in that situation, is also located away from the driving source. However, the resulting behavior is much richer, as we have broken the left-right symmetry by choosing different strengths of $K_1$ and $k_2$ and $m_1$ and $m_2$. Translating these considerations into the description of spin-lasers in Fig.~\ref{fig:cho9}, 
for the IM response of $S_x$ at the lower resonance frequency 
$f_L$, we see an example of Fano-type resonance~\cite{Miroshnichenko2010:RMP,Limonov2017:NP}.

Since the model of driven coupled harmonic oscillators and their phase relations offers a classical picture for the origin of the Fano resonance~\cite{Joe2006:PS}, we can also apply it to gain insights into spin-lasers where the oscillator $S_x$ is pulled and pushed by the modulation source and the oscillator $S_y$.  
At $f\gtrsim f_L$, the two forces almost cancel because $S_y$ is out of phase with the modulation source and as a consequence, there is a minimum, the antiresonance, which is a characteristic feature of 
Fano interference~\cite{Joe2006:PS}. 
The interference of the broad, higher-resonance peak, representing the continuum of states and the narrow lower resonance peak,
representing the discrete state,
produces an asymmetrical line shape $\sigma(\varepsilon)$ 
characterizing absorption spectra, scattering cross-section, transmission coefficient, and other phenomena
~\cite{Connerade1988:RPP,Limonov2017:NP,Miroshnichenko2010:RMP} 
\begin{equation}
\sigma(\varepsilon) = D^2\left(q + \varepsilon \right)^2/\left(1 + \varepsilon^2 \right),
\label{eq:Fano}
\end{equation}
where $\varepsilon = 2\left( f - f_L \right)/\Delta f_L$, 
$\Delta f_L$ is the resonance width, and from the line shape fit 
we obtain $D^2 = 0.35$, $\Delta f_L = 1.34\ \mathrm{GHz}$, and the Fano parameter $q = - 3.25$~\cite{Limonov2017:NP}. In the limit $q \rightarrow \pm \infty$, there is no asymmetry, and Eq.~(\ref{eq:Fano}) recovers the Lorentizan shape~\cite{Connerade1988:RPP}.

While Fano resonances and the reflection of light on one of the mirrors in the resonant cavity are an integral part of some
Fano lasers, offering a path toward
energy-efficient and ultrafast operation~\cite{Deng2021:AOM}, what 
we find in Fig.~\ref{fig:cho9} is another manifestation of 
Fano resonance in lasers. The suppression of the one intensity component (in our case the $x$ component) in the narrow frequency interval around antiresonance, as expected for Fano resonance, 
is the property of the gain region itself rather than the previously considered specific property of the mirror~\cite{Deng2021:AOM}. 
This is just one of the examples where the model of coupled harmonic oscillators could guide the understanding
of unexplored regimes in spin-lasers. 

By establishing the connection between the model of coupled harmonic
oscillators and the direct modulation of spin-lasers, we have mostly focused on recovering similar main trends between the two systems to further elucidate the operation of spin-lasers. 
In this approach, we have not tried to explore a large parameter range for spin-lasers
or their different experimental implementations, both deserving detailed separate studies. 
For example, our typical choice of the injected steady-state polarization was just 4\%. This value can be readily achieved and exceeded by optical and electrical injection, while the corresponding circular polarization of the emitted light can be also realized using the chiral 
properties of metasurfaces and cavities~\cite{Jia2023:O,Maksimov2022:PRA}. 

We have considered short spin-relaxation times in 
quantum well
III-V semiconductors $\tau_s=2\!\!-\!\! 20\ \mathrm{ps}$, which can be made even shorter~\cite{Lindemann2019:N} or much longer, exceeding nanoseconds 
at $300\ \mathrm{K}$~\cite{Iba2011:APL,Iba2021:M}. 
However, in other systems the hole spin-relaxation time may not be negligible and could be comparable with the one for electrons. This was discussed for wurtzite GaN-based lasers with modest spin-orbit coupling, where the rate equations were combined with the microscopic gain description to yield the lasing threshold with nonmonotonic dependence on electron-spin polarization~\cite{FariaJunior2017:PRB}.  From these steady-state results, we can infer that the IM response could also lead to nonmonotonic $f_\text{3dB}$ in $P_{J0}$, which is different from what was obtained from the SFM or intensity equations~\cite{Xu2021:PRB}, unless they are generalized to include spin-polarized holes. 

Another realization of comparable spin-relaxation times for electrons and holes is in III-V quantum dots (QDs)~\cite{Basu2008:APL}. A theory of QD spin-lasers has shown the need to generalize the rate equations~\cite{Gothgen2008:APL} to include both spin-polarized electrons and holes, the presence of the wetting layer, as well as carrier occupancies with Pauli blocking factors~ \cite{Oszwaldowski2010:PRB}. A different QD implementation was derived from topological insulators (TIs) in the inverted HgTe/CdTe quantum wells, which can support many optical transitions~\cite{Scharf2015:PRB}. The obtained rate equations, for optical~\cite{Huang2019:PRA} or electrical carrier injection~\cite{Huang2021:SCPAM} provide valuable insights for THz emission using the helical edge states in TI QDs. While these studies were focused on the steady-state operation and excluded birefringence, the optical selection rules with $\Delta J_z=\pm 1$ are the same as used in the SFM~\cite{SanMiguel1995:PRA,Lindemann2019:N}, this work, and the previous rate equations~\cite{Gothgen2008:APL}. The levels included for the stimulated spin-flip transitions in TI QDs resemble the four-level description of the SFM and intensity equations~\cite{SanMiguel1995:PRA,Xu2021:PRB}. Since quantum well-based lasers can be mapped to QD lasers~\cite{Lee2012:PRB}, it would be interesting to explore if the model of coupled harmonic oscillators could also guide various QD-based lasers.

One of the assumptions in our work is the presence of 
a large birefringence that could push the modulation response of spin-lasers 
beyond the frequencies attained by their conventional counterparts. There is a growing number of approaches to implement such a birefringence, not only using 
elasto-optical effect, strain, thermal gradients, and surface 
gratings~\cite{Panajotov2000:APL,FariaJunior2015:PRB,Pusch2017:APL,Pusch2019:EL,Pusch2015:EL,Lindemann2023:EL}
but also with photonic crystals, columnar thin films, 
and nematic liquid crystals~\cite{Dems2008:OC,Panajotov2024:OL,Lempicka-Mirek2023:N}. Another assumption is the implementation of fast PM. In addition to the previous well-established optical approaches, there are encouraging
advances toward electrical implementation of PM in other systems. With the recent breakthroughs of using magnetization dynamics from spin-orbit torque to electrically modulate the helicity of the emitted photons in light-emitted diodes at zero applied magnetic field and 
$300\ \mathrm{K}$~\cite{Dainone2024:N,Hiura2024:N}, 
there is a push to transfer this principle to spin-lasers, where the current modulation is limited to the optical spin injection. However, there are many other opportunities to modulate the spin population in the gain region of lasers, including in
different material systems~\cite{Battiato2010:PRL,Kampfrath2013:NN,Kirilyuk:2019,%
vandenBrink2016:NC,Hendriks2024:NC,Zutic2019:MT,Rozhansky2020:PRB,%
Rozhansky2023:JMMM,Safarov2022:PRL,Xu2020:PRL,Nishizawa2017:PNAS,%
Nishizawa2018:APE,Munekata2020:SPIE}. 

Building on these recent advances and a growing interest in spintronics beyond
magnetoresistance, 
we are encouraged that our findings will be relevant to other paths toward implementing spin-lasers and that they can elucidate how 
such devices could exceed the performance of their classical counterparts. 
Furthermore, our studies may stimulate additional investigation of coupled  
oscillators  as model systems to study dynamical properties of other 
spintronic devices~\cite{Dery2011:APL,Khaetskii2013:PRL,Stadler2014:PRL,Hirohata2020:JMMM,Tsymbal:2019}. The phase evolution, important in lasers and emerging logic devices~\cite{Csaba2020:APR}, also offers motivation to generalize models
of coupled oscillators and implement unexplored modulation schemes, as can be seen from modeling Josephson junctions, which are described by a driven pendulum, rather than by a driven harmonic oscillator~\cite{Tafuri:2019,Monroe2022:PRA,Monroe2024:APL}.

\section*{acknowledgments}
This work has been supported by National Science Foundation Grants No. ECCS-2512491 and No. ECCS-2130845. 
We thank N. C. Gerhardt and Y. Lu for valuable discussions.

\appendix

\section{\label{sec:app}Appendix}
While the employed analogy of coupled harmonic oscillators with spin-lasers is not pursued here to obtain quantitative agreement, 
we can accurately recover various main trends. A simple understanding of the coupled harmonic oscillators tells us what to expect if CASE 1 is characterized by a single resonance at the lower angular frequency $\omega_L(k)$, while CASE 2 has two resonances at 
$\omega_L(k)$ and at a higher value $\omega_H(k)$, recall Eqs.~(\ref{eq:omegaLfull}) and (\ref{eq:omegaHfull}). Specifically, at 
$1>|\mu|>0$, the ratio between the amplitudes of the displacement due to external forces in CASE 1 and CASE 2, the two-peak response should appear, even if it was absent for $|\mu|=1$ in Fig.~\ref{fig:cho4}(a). Indeed, we see in Fig.~\ref{fig:cho10}(a) that, by reducing $|\mu|$, we can gradually recover the two-peak response and an enhanced bandwidth from CASE 2 in Fig.~\ref{fig:cho3}(a) with $|\mu|=0$.

\begin{figure}[t]
\centering
\includegraphics*[width=8.6cm]{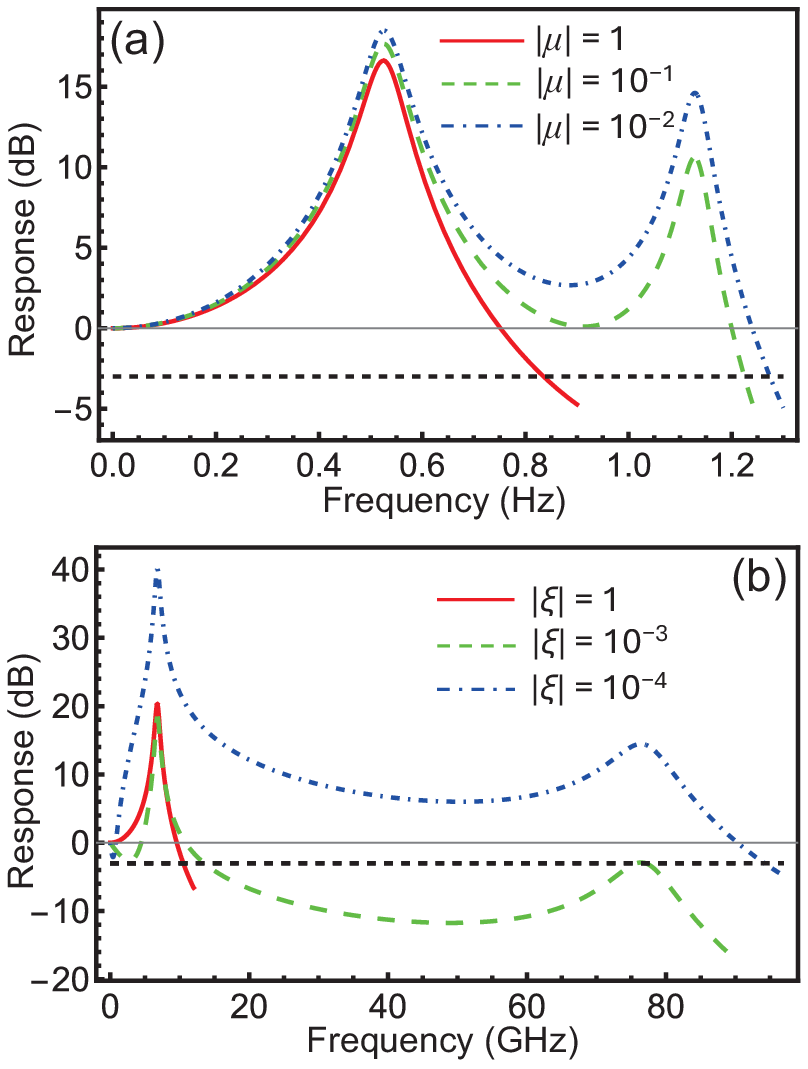}
\vspace{-0.5cm}
\caption{The evolution of the response functions from Fig.~\ref{fig:cho4} for different amplitude ratios in the combined modulation cases,
for coupled harmonic oscillators $|\mu|$, and for spin-lasers $|\xi|$. (a) Response function for $x_2$ with a combined modulation of the coupled harmonic oscillators from Fig.~\ref{fig:cho1}(c) described by 
$|\mu| =1$, identical to Fig.~\ref{fig:cho4}(a), 
$|\mu| =0.1$ and 0.01. (b) 
Response function for $S_y$ with a simultaneous IM and PM for a spin-laser, 
described by 
$|\xi| =1$, identical to Fig.~\ref{fig:cho4}(b), 
$|\xi| =10^{-3}$ and $10^{-4}$. 
The phases of $\mu$ and $\xi$ are both $\pi/3$.}
\label{fig:cho10}
\vspace{-0.3cm}
\end{figure}

By applying this guidance to spin-lasers, we expect similar trends and the recovery of the second resonance peak, at $\omega_H$ given by 
Eq.~(\ref{eq:omegaH2}),
for a small enough $|\xi|= |\delta J_0/\delta P_{J0}|$, which describes the ratio of the amplitudes for IM and PM. This expectation from the behavior of the coupled harmonic oscillators is also verified in 
Fig.~\ref{fig:cho10}(b). We see that by reducing $|\xi|=1$, considered also 
in Fig.~\ref{fig:cho4}(b) we can gradually recover exactly the two-peak response and an enhanced bandwidth from PM in Fig.~\ref{fig:cho3}(b) with $|\xi|=0$. However, the value of $|\xi|=10^{-4}$, where the high-resonance response exceeds the signal-to-noise threshold of  
$f_\text{3dB}$ (dashed line) is $\sim \!3$ orders of magnitude smaller than the corresponding value of $|\mu|$.

\bibliographystyle{apsrev4-2}

\end{document}